\documentclass[lineno]{jfm}
\usepackage{graphicx}
\usepackage{epstopdf,epsfig}
\usepackage[justification=justified,singlelinecheck=false]{caption}
\usepackage{dirtytalk}
\usepackage{newtxtext}
\usepackage{gensymb}
\usepackage{newtxmath}
\usepackage{natbib}
\usepackage{tabularx}
\usepackage{pdflscape}
\usepackage{booktabs, makecell}
\usepackage{rotating}
\usepackage{hyperref}
\usepackage{caption}
\usepackage{subfigure}
\usepackage{subcaption}
\usepackage [autostyle, english = american]{csquotes}
\MakeOuterQuote{"}
\hypersetup{
colorlinks = true,
urlcolor   = blue,
citecolor  = black,}

\newcommand{\RomanNumeralCaps}[1]
\linenumbers

\shorttitle{Journal of Fluid Mechanics}
\shortauthor{P. Patra, S. Warrier and A. Roy}

\title{The electrostatic–hydrodynamic interaction of two spheres in a linear compressional flow}
\author{Pijush Patra \footnote{Present address: Nordita, KTH Royal Institute of Technology and Stockholm University, Stockholm 10691, Sweden; Email:  pijush.patra@su.se}, Srikumar Warrier
 \and Anubhab Roy
 \corresp{\email{anubhab@iitm.ac.in}}}
\affiliation{Department of Applied Mechanics and Biomedical Engineering, Indian Institute of Technology Madras, Chennai, Tamil Nadu 600036, India.}

\begin{document}
\maketitle

\begin{abstract}

We investigate the collision dynamics of two like-charged dielectric spheres in a uniaxial compressional flow while accounting for non-continuum lubrication interactions, van der Waals attraction, and the complete finite-dielectric electrostatic interaction. Mutual polarization causes the electrostatic force to behave substantially differently from the point-charge approximation and can produce strong near-field attraction despite far-field Coulomb repulsion, depending on the particle size and charge ratios. We identify a critical electrostatic strength above which particles approaching from the far field cannot reach contact. This threshold corresponds to a saddle-node bifurcation of the relative-trajectory field on the compressional axis. Below this threshold, near-field polarization-induced attraction can increase the upstream interception area and substantially enhance the collision efficiency. We report that electrostatic interactions lead to non-monotonic variations of collision efficiency with electrostatic strength, particle size ratio, and charge ratio. When electrostatic interactions remain repulsive at all separations, increasing their strength suppresses collisions monotonically. Non-continuum lubrication and van der Waals interactions further modify the collision efficiency, although sufficiently strong polarization-induced attraction can dominate these near-field effects. For weak electrostatic interactions, a perturbation analysis expresses the leading correction to the collision efficiency as a weighted average of the electrostatic force along the limiting colliding trajectory, with the weight concentrated in the lubrication region; this prediction agrees closely with the numerical results. These results demonstrate that finite-size polarization and near-field hydrodynamic interactions jointly govern the collision behaviour of like-charged particles in compressional flows.

\end{abstract}


\section{Introduction} \label{sec:Introduction}

The transport and collision of particles suspended in a gaseous medium are central to a wide range of atmospheric and geophysical processes, including droplet growth in clouds and particle aggregation in volcanic plumes. In such systems, the relative motion of particle pairs is governed by background flow and gravity, while hydrodynamic and non-hydrodynamic interactions can substantially modify their motions and collision outcomes. Of particular interest are electrostatic interactions, since cloud droplets and volcanic ash particles can acquire charge through several mechanisms, including ion attachment, inductive charging, and triboelectrification \citep{pruppacher1997microphysics,harrison2004global,mendez2016effects}. The dominant charging mechanisms, however, differ between these environments. Droplets in warm clouds can acquire charge primarily through the attachment and field-dependent collection of atmospheric ions \citep{harrison2003ion}, whereas volcanic ash is commonly electrified through particle-particle contact (triboelectrification) and fragmentation-related processes. The latter can additionally depend on environmental conditions such as temperature and water content \citep{mendez2020microphysical}. The electrostatic forces arising from surface charges can either promote or inhibit particle encounters, depending on particle size and charge, thereby influencing collision-coalescence of water droplets in clouds and the aggregation of ash particles in volcanic plumes. Electrification is particularly evident in convective clouds and ash-rich volcanic plumes, where intense electrical activity and lightning provide striking manifestations of particle charging \citep{wang2013physics,behnke2018impulsive,ichihara2023multiphase,reeve2025electrostatic}. These observations motivate a fundamental fluid-mechanical question: how does the coupling between electrostatic forces and hydrodynamic interactions modify the relative motions and collision dynamics of finite-sized particles in a background flow?

The fluid-mechanical framework for understanding particle collisions in turbulent flows has its origins in the seminal work of \citet{saffman1956collision}, who considered particles much smaller than the Kolmogorov length scale and approximated the turbulent velocity field experienced by an approaching pair as a quasi-steady uniaxial compressional flow. For sub-Kolmogorov particles, the flow varies linearly over the interparticle separation, and the instantaneous relative motion induced by turbulence can therefore be described in terms of the local velocity-gradient tensor. The straining component of this tensor drives particles towards one another, thereby serving as a crucial mechanism for collisions. Using Gaussian statistics for turbulent strain rates, \citet{saffman1956collision} obtained the collision rate to be $(8 \pi /15)^{1/2} n_1 n_2 \varGamma_{\eta} (a_1+a_2)^3$ where $a_1$ and $a_2$ are the particle radii, $n_1$ and $n_2$ are their respective number densities, and $\varGamma_{\eta}=(\epsilon/\nu_f)^{1/2}$ is the Kolmogorov shear rate, with $\epsilon$ and $\nu_f$ denoting the turbulent dissipation rate and the kinematic viscosity of the surrounding fluid, respectively. \citet{zeichner1977use} calculated the collision rate of two non-interacting spheres subjected to a steady uniaxial compressional flow to be $(8 \pi /3\sqrt{3})n_1 n_2 \dot{\gamma} (a_1+a_2)^3$, where $\dot{\gamma}$ is the compression rate. Equating this expression with the Saffman–Turner collision rate yields $\dot{\gamma}=(9/40\pi)^{1/2}\varGamma_{\eta}$, thereby relating the prescribed compressional flow to the characteristic small-scale strain rate of turbulence. The consideration of uniaxial compressional flow is particularly salient, as the most probable local strain in turbulence has two extensional and one compressional principal direction \citep{ashurst1987alignment}, a state whose axisymmetric idealization is the uniaxial compressional flow considered here. This characteristic motivates us to perform a focused analysis of the mechanics underlying particle-pair interactions in a canonical flow configuration. In the present investigation, we adopt the assumption of a frozen uniaxial compressional flow as the background flow field.

A local linear flow approximation does not fully capture the complexities of turbulence dynamics. Notably, turbulence and particle inertia can influence the collision rate via (i) altering the radial relative velocity of particle pairs and (ii) preferential concentration \citep{sundaram1997collision,chun2005clustering,falkovich2007sling}. Consequently, a collision kernel obtained by approximating the background flow as a steady compressional flow should not be interpreted as the turbulent collision kernel. Previous studies on charged particles in turbulence show that electrostatic effects remain weak when Coulomb interactions are small compared with turbulence-induced relative motion, but become important once the two are comparable \citep{lu2015charged,ruan2024effects}. Furthermore, since the Coulomb force increases rapidly as particle separation decreases, electrostatics can play an important role during the final approach of particles, even when turbulence controls the larger-scale encounter dynamics. These studies, however, considered point-charge Coulomb interactions and did not include the finite-size polarization effects examined here.
The trajectory-based formulation employed here addresses a more specific fluid-mechanical problem: it quantifies how hydrodynamic interactions and non-hydrodynamic forces modify pair trajectories and the collision dynamics for a prescribed straining flow. This formulation allows us to isolate the coupling between electrostatic and hydrodynamic interactions without conflating it with turbulence-induced clustering and relative-velocity statistics. Nevertheless, researchers have extended trajectory-based approaches to turbulent flows by approximating the background turbulent flow in the vicinity of an inertialess particle pair as a stochastically varying linear flow \citep{brunk1998turbulent,chun2005coagulation,dhanasekaran2021tubulent}.

When two particles approach one another, the disturbance flow generated by each particle modifies the flow field experienced by the other, giving rise to hydrodynamic interactions that become increasingly important at small interparticle separations. These interactions can substantially modify pair trajectories and, consequently, influence collision rates. The effect of these hydrodynamic interactions can be quantified through the collision efficiency, defined as the ratio of the collision rate in the presence of hydrodynamic interactions to that in their absence. Two-body hydrodynamic interactions under Stokes-flow conditions have been studied extensively \citep{guazzelli2011physical,kim2013microhydrodynamics}. In the case of rigid spheres, continuum lubrication resistance to normal relative motion scales as $O(1/\xi)$, thereby preventing contact in finite time. Conversely, for viscous drops, interfacial mobility weakens the lubrication resistance, leading to an $O(1/\sqrt{\xi})$ scaling \citep{davis1989lubrication}. Here, $\xi$ is the closest surface-to-surface separation distance normalized by the mean radius of the two particles. The role of continuum hydrodynamic interactions in the collisions of rigid particles or viscous drops within uniaxial compressional flow has been examined previously \citep{zeichner1977use,wang1994collision}. 

For particles suspended in a gaseous medium, like air, the continuum approximation becomes invalid when the interparticle separation is less than the molecular mean free path. Under such situations, one needs to consider non-continuum lubrication resistance, which scales as $O(\ln(\ln(Kn/\xi))/Kn)$ \citep{sundar96non}. This non-continuum lubrication resistance exhibits a weaker divergence than its continuum counterpart, thereby allowing collisions in finite time. Here, the Knudsen number $Kn$, which measures the strength of non-continuum interaction, is defined as the ratio of the mean free path of the medium to the mean radius of the particles. Previous studies have quantified the influence of non-continuum hydrodynamic interactions on collision dynamics for particles undergoing thermal fluctuations \citep{patra2022brownian}, suspended in laminar flows \citep{dhanasekaran2021collision,patra2022collision}, and turbulent flow \citep{dhanasekaran2021tubulent,dhanasekaran2025effect}.

To assess the influence of surface charge on collision dynamics, both the magnitude and the polarity of the particle charges must be specified. Field measurements have established empirical relations between droplet charge and size (see Chap.~18.4 of \citealt{pruppacher1997microphysics}; \citealt{beard2004aircraft,zhou2007production,nicoll2016stratiform}), and observed cloud-droplet charges lie well below the Rayleigh limit, beyond which electrostatic stresses destabilize the interface \citep{rayleigh1882xx,pruppacher1997microphysics}. Even such subcritical charges can modify collision dynamics, because mutual polarization redistributes surface charge during close approach and the interaction then departs markedly from the point-charge Coulomb force.

The electrostatic interaction between two charged spheres has been studied extensively, beginning with the inverse-distance expansion of \citet{maxwell1873treatise}, followed by the capacitance formulation of \citet{russell1909coefficients} and the bispherical-coordinate solution for conducting spheres developed by \citet{davis1964two}. The far- and near-field asymptotic behaviour of the electrostatic force between perfectly conducting spheres is also well established \citep{lekner2012electrostatics,lekner2016regions}. A notable consequence of charge redistribution is that two like-charged conducting spheres may experience an attractive interaction at sufficiently small separations, except for charge ratios corresponding to equal surface potentials at contact. In the near-contact limit, this attraction exhibits an $O\left(\xi^{-1}[\ln\xi]^{-2}\right)$ singularity \citep{lekner2012electrostatics}, which is sufficiently strong to overcome continuum lubrication resistance and allow finite-time contact.

The perfect-conductor approximation is sufficiently accurate for particles with very high dielectric constants, such as metallic particles. However, this approximation is not appropriate for many common dielectric materials, including water, which has a relative permittivity of approximately $80$. Consequently, cloud droplets may exhibit an electrostatic response that differs appreciably from that of perfect conductors, particularly at small interdroplet separations. Previous studies have shown that, depending on the particle size and charge ratios, two like-charged dielectric spheres can experience an attractive electrostatic interaction at sufficiently small surface-to-surface separations \citep{bichoutskaia2010electrostatic,munirov2013interaction,khachatourian2014electrostatic}. The physical origin of this behaviour lies in mutual polarization. At large separations, the leading electrostatic interaction is the monopole--monopole Coulomb contribution, which is repulsive for like-charged particles. As the particles approach, however, each particle increasingly polarizes the other, generating a spatially non-uniform bound-charge distribution. The resulting higher-order polarization contributions become progressively stronger and can overcome the Coulomb repulsion, causing the net electrostatic force to change from repulsive at large separation to attractive at close separation.

We define the size ratio $\kappa=a_2/a_1$ and the charge ratio $\beta=q_2/q_1$, where $q_1$ and $q_2$ are the charges carried by the larger and smaller spheres of radii $a_1$ and $a_2$, respectively. For finite dielectric permittivity, the near-contact interaction is repulsive only within a finite band of $\beta$ at each $\kappa$, outside which polarization-induced attraction prevails; in the perfect-conductor limit this band collapses to a single curve \citep{patra2023collision}. Although the interaction between like-charged dielectric spheres approaches that between conducting spheres at large separations, the two differ substantially in the lubrication regime, where polarization-induced attraction and hydrodynamic resistance both become strong and compete directly in determining whether an approaching pair reaches contact.

The pairwise collision dynamics of colloidal particles in linear flows, incorporating the combined effects of hydrodynamic and colloidal interactions, including screened double-layer repulsion, have been studied extensively in classical colloid science \citep{zeichner1977use,russel1989}. In a gas, by contrast, electrostatic interactions are unscreened and long-ranged, and mutual polarization becomes important at close approach. Most previous studies of electrostatically modulated collisions in gases have considered conducting spheres settling under gravity in a quiescent fluid \citep{schlamp1976numerical,schlamp1979numerical,klimin1994collision,khain2004rain,guo2021enhancement,dubey2024critical,zhang2025collision}, and, with few exceptions \citep{dubey2024critical}, they have relied on simplified descriptions of the hydrodynamic or electrostatic interactions that do not resolve their coupling in the lubrication regime. \citet{patra2023collision} resolved this coupling for like-charged dielectric spheres settling under gravity. In an imposed linear flow, electrostatic effects on collisions in a gas have, to our knowledge, been examined with full hydrodynamic interactions only for uncharged droplets polarized by an external electric field \citep{patra2025electric}, the settling counterpart of which was studied by \citet{patra2025gravity}. The interplay between non-continuum lubrication and the complete electrostatic interaction between charged, finite-dielectric particles in an imposed linear flow has not been addressed. This study aims to address this gap by combining non-continuum hydrodynamic interactions with the complete finite-dielectric electrostatic interaction to quantify the effects of mutual polarization on pair trajectory and collision efficiency of finite-sized charged particles in a uniaxial compressional flow.

The remainder of the paper is organized as follows. In \S \ref{sec:Problem_formulation}, we present the problem formulation and the framework used to evaluate the collision rate and efficiency. We briefly describe the numerical method in \S \ref{Numerical_method}. Then, in \S \ref{sec: trajectory calculations}, we show pair trajectories and the collision efficiency arising from the interplay between hydrodynamic and electrostatic interactions. Finally, in \S \ref{sec: Conclusion}, we summarize the principal findings.

\section{Problem formulation} \label{sec:Problem_formulation}
\subsection{Trajectory equations for a particle pair} \label{Trajectory_equations_for_a_particle_pair}

\begin{figure}
    \centering
    \includegraphics[width=0.8\linewidth]{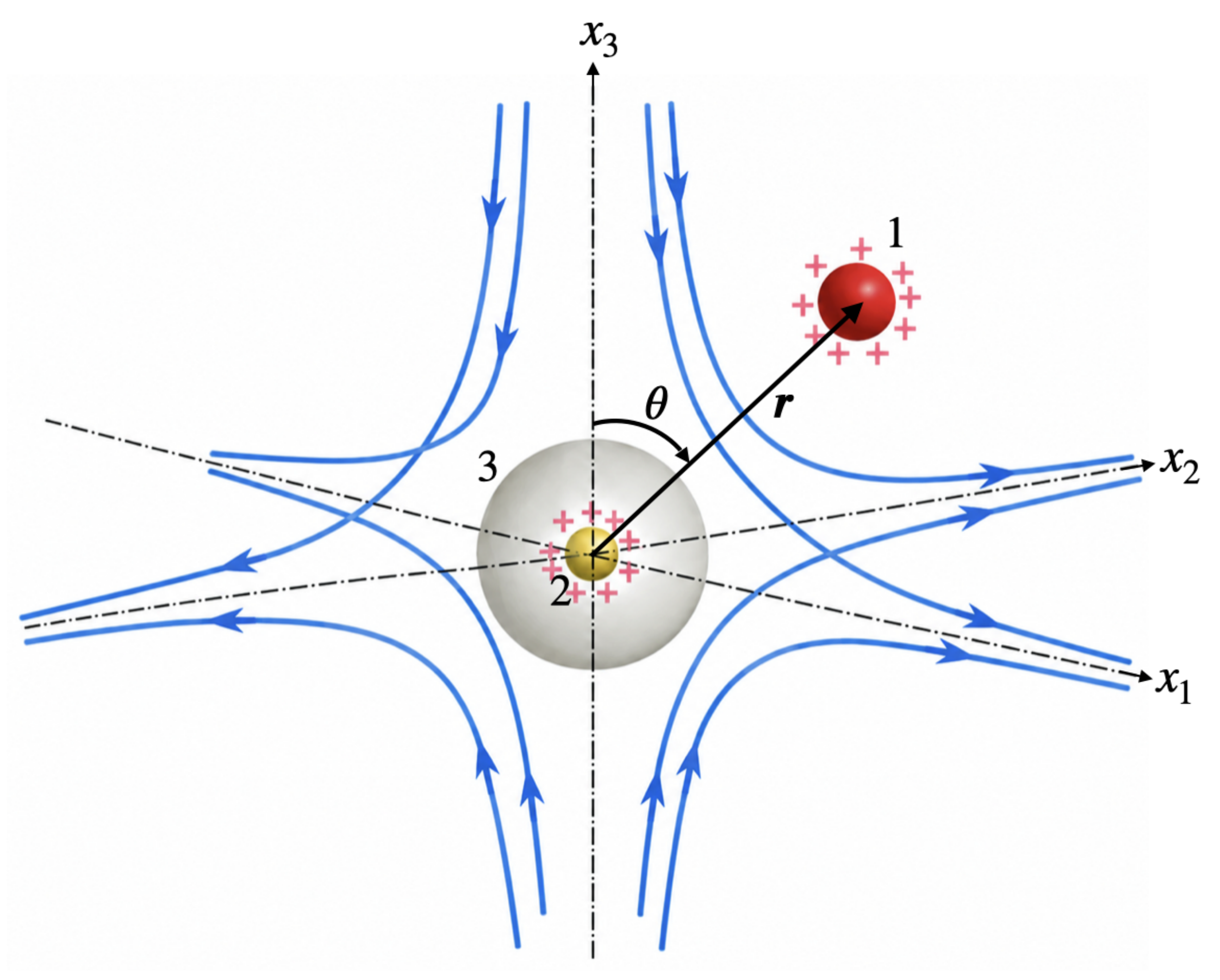}
    \caption{Schematic of two like-charged particles of radii $a_1$ and $a_2$ in a uniaxial compressional flow. Here, $\boldsymbol{r}$ is the centre-to-centre separation vector, and $\theta$ is the angle between the compressional axis (i.e., $x_3$-axis) and $\boldsymbol{r}$. Sphere `1' denotes the satellite sphere and sphere `2' the test sphere, while sphere `3' represents the collision sphere of radius $a_1+a_2$, centred on the test sphere. Representative trajectories of the satellite sphere relative to the test sphere are shown in blue. We denote the unit vectors in the $r$ and $\theta$ directions as $\hat{\boldsymbol{e}}_{r}$ and $\hat{\boldsymbol{e}}_{\theta}$, respectively.}
    \label{Schematic}
\end{figure}

We consider a dilute polydisperse suspension of charged spheres subjected to a uniaxial compressional flow described by the velocity field $\textit{\textbf{U}}^\infty(\boldsymbol{x}) = (\dot{\gamma}x_1,\dot{\gamma}x_2,-2\dot{\gamma}x_3)$, where $\dot{\gamma}$ is the compression rate, $x_1$ and $x_2$ correspond to the extensional axes, and $x_3$ corresponds to the compressional axis. In many natural and industrial systems, such as atmospheric clouds and aerosol reactors, particle volume fractions are typically very low, about $O(10^{-6})$ (see \citealt{grabowski2013growth,balthasar2002detailed}). This allows us to treat these systems as dilute, in which the likelihood that a third particle affects the relative motion between two interacting particles is negligible. Therefore, our focus will be on the binary interactions of two spherical particles with radii $a_1$ and $a_2$ (see Figure \ref{Schematic}).

Before proceeding with the detailed analysis, it is essential to justify the key assumptions made in this study. The present investigation does not consider the role of surrounding fluid inertia. While particle inertia can significantly influence collision dynamics for larger particles, this effect is negligible for relatively small particles. Additionally, we assume that the particles are sufficiently large, so Brownian diffusion is negligible. The particle Reynolds number, denoted as $Re_p$, is defined in terms of the compression rate $\dot{\gamma}$ and the radius of the larger particle $a_1$ as follows: $Re_p=\rho_f\dot{\gamma} a_1^2/\mu_f$, where $\rho_f$ and $\mu_f$ are the density and dynamic viscosity of the surrounding fluid, respectively. The Stokes number, denoted by $St$, captures the effects of particle inertia and is defined as: $St=\dot{\gamma}\tau_p$, with $\tau_p=2 a_1^2\rho_p/(9\mu_f)$ representing the viscous relaxation time of the larger particle, where $\rho_p$ is the particle density. The P\'eclet number, denoted by $Pe$, measures the relative strengths of flow advection and Brownian diffusion and is given by: $Pe=3\pi \mu_f \dot{\gamma}a_1^3/k_BT$. Here, $k_B=1.381 \times 10^{-23}$JK$^{-1}$ is the Boltzmann constant, and $T$ is the absolute temperature. Let us calculate these relevant dimensionless quantities for a water droplet in air with $a_1=15$ \textmu m, $\rho_p \approx 10^3$ kg m$^{-3}$, $\rho_f \approx 1$ kg m$^{-3}$, $\mu_f \approx 1.8 \times 10^{-5}$ Pa s, $T=298$ K, and $\dot{\gamma}=25$ s$^{-1}$ corresponding to $\varepsilon \approx 0.16$ m$^2$ s$^{-3}$, representative of vigorous convective clouds. Using this data, we obtain the following estimates: $Re_p \approx 3.125 \times 10^{-4}$, $St \approx 0.069$ and $Pe \approx 3.478 \times 10^3$. These typical values of $Re_p$, $St$, and $Pe$ substantiate the assumptions made in the present study.

The characteristic hydrodynamic and capillary stresses for a spherical water droplet of radius $a_1$ in a uniaxial compressional flow scale as $\mu_f\dot{\gamma}$ and $\Upsilon/a_1$, respectively, where $\Upsilon$ is the surface tension of the droplet-air interface, giving the flow capillary number $Ca=\mu_f\dot{\gamma}a_1/\Upsilon$. For $a_1=15$ \textmu m, $\dot{\gamma}=25\,\mathrm{s}^{-1}$ and $\Upsilon = 0.073$ Nm$^{-1}$, $Ca\approx9.2\times10^{-8}$, indicating negligible flow-induced deformation. Moreover, for water droplets in typical warm-cloud conditions, the high droplet-to-medium viscosity ratio ($\approx 10^2$) suppresses interfacial mobility and internal circulation, thereby justifying the approximation of small cloud droplets as rigid spheres. Consistent with this approximation, \citet{dhanasekaran2021collision} showed that non-continuum lubrication and van der Waals interactions dominate over interfacial mobility in determining droplet collisions (see their figure 2).

Gravitational settling is not included in the present formulation. Its importance relative to the imposed strain is measured by the settling parameter $S_v = 4\rho_p g\,(a_1 - a_2)/(9\mu_f\dot{\gamma})$, where $g$ is the gravitational acceleration; $S_v$ is the ratio of the difference in Stokes settling velocities of the two particles to the strain-induced relative velocity $\dot{\gamma}(a_1+a_2)/2$. For water droplets with $a_1 = 10\,\upmu\mathrm{m}$ and $\dot{\gamma} = 25\,\mathrm{s}^{-1}$, $S_v \approx 10$ at $\kappa = 0.9$ and $S_v \approx 48$ at $\kappa = 0.5$, whereas $S_v = 0$ for equal-sized droplets. Differential settling therefore dominates the relative motion of strongly bidisperse cloud droplets at these strain rates. The present results describe the strain-dominated limit $S_v \lesssim O(1)$, which applies to near-equal pairs, for which straining is the only mechanism of approach, and to strongly strained environments. The coupled problem of settling in a compressional flow has been solved for uncharged spheres by \citet{dhanasekaran2021collision}.

When $Re_p \ll 1$, the fluid motion around the particle pair can be accurately described using the Stokes equations. In the limit where both $Re_p$ and $St$ are zero, we can express the relative velocity between two particles as a linear superposition of their relative velocities induced by the background flow, electrostatic forces, and van der Waals forces. For the current problem, the relative velocity between the two spheres, denoted by $\boldsymbol{V}_{12}$, is given as:
\begin{eqnarray}
    \boldsymbol{V}_{12} = \boldsymbol{V}_{1} - \boldsymbol{V}_{2} =\boldsymbol{E^{\infty}} \boldsymbol{\cdot} \boldsymbol{r}- \Big[A\frac{\boldsymbol{r}\boldsymbol{r}}{r^{2}}+B\left(\boldsymbol{I}-\frac{\boldsymbol{r}\boldsymbol{r}}{r^{2}}\right)\Big] \boldsymbol{\cdot} \left(\boldsymbol{E^{\infty}} \boldsymbol{\cdot} \boldsymbol{r}\right) \nonumber \\ + \frac{1}{6\pi\mu_{f}}\left(\frac{1}{a_{1}}+\frac{1}{a_{2}}\right)\left[G\frac{\boldsymbol{r}\boldsymbol{r}}{r^{2}}+H\left(\boldsymbol{I}-\frac{\boldsymbol{r}\boldsymbol{r}}{r^{2}}\right)\right] \boldsymbol{\cdot} \left(\boldsymbol{F}_e + \boldsymbol{F}_{\text{vdW}}\right),
\label{Eq: relative_velocity} 
\end{eqnarray}
where $\boldsymbol{V}_{1}$ and $\boldsymbol{V}_{2}$ denote the velocities of the satellite sphere (with radius $a_1$) and the test sphere (with radius $a_2$), respectively, $\textit{\textbf{E}}^{\infty} =  [(\boldsymbol{\nabla} \textit{\textbf{U}}^{\infty}) + (\boldsymbol{\nabla} \textit{\textbf{U}}^{\infty})^T]/2$ represents the strain rate tensor, $\boldsymbol{r}$ is the separation vector that points from the center of the test sphere to the center of the satellite sphere, $r = |\boldsymbol{r}|$, $\textbf{\textit{I}}$ is the unit second-order tensor, and $\boldsymbol{F}_e$ and $\boldsymbol{F}_{\text{vdW}}$ represent the electrostatic and van der Waals forces, respectively. The mobility functions $A$, $B$, $G$, and $H$ capture the hydrodynamic interactions. Specifically, $A$ and $G$ are axisymmetric mobilities, while $B$ and $H$ are asymmetric mobilities corresponding to linear flow and non-hydrodynamic forces, respectively. For continuum hydrodynamic interactions, these mobilities depend on the size ratio $\kappa = a_2/a_1$, and the dimensionless center-to-center distance $r/a^*$ of the particle pair, where $a^*=(a_1+a_2)/2$ represents the average radius of the two spherical particles. The dimensionless surface-to-surface distance $\xi$, which we have used in the previous section, can be expressed as $\xi = (r-(a_1+a_2))/a^* = (r/a^*)-2$. Methods for computing the continuum hydrodynamic mobilities, and their asymptotic forms at small and large separations, are well documented in the literature \citep{batchelor1972hydrodynamic,kim1985resistance,jeffrey1992calculation,wang1994collision}.

The continuum assumption for hydrodynamic interactions would not be valid for particles colliding in a gaseous medium, particularly when the separation distance between the surfaces of the two spheres is less than the mean free path of the medium, denoted as $\lambda_0$. The mean free path of air increases with altitude and is approximately 0.1 \textmu m under typical warm cumulus cloud conditions \citep{wallace2006atmospheric}. Since we consider scenarios where suspending media are gases, we must account for the non-continuum lubrication interactions. In these cases, hydrodynamic mobilities additionally depend on the Knudsen number, defined as $Kn=\lambda_0/a^*$. The expressions of non-continuum lubrication forces for axisymmetric motion have been derived by \citet{sundar96non}. Building upon this foundational work, \citet{dhanasekaran2021collision} have recently incorporated the non-continuum lubrication interactions for axisymmetric mobilities. They considered continuum lubrication interactions for $\xi > O(Kn)$ and non-continuum lubrication interactions for $\xi \leq O(Kn)$. In the current study, we will utilize the uniformly valid solutions they developed for $A$ and $G$. Continuum asymmetric mobilities $B$ and $H$ approach a finite value as $\xi \rightarrow 0$. Consequently, we anticipate that the breakdown of the continuum assumption will exert a negligible influence on the asymmetric relative motions of an inertialess particle pair. Therefore, we will consider continuum hydrodynamics for asymmetric mobilities at all separations.

The van der Waals attraction force, $\boldsymbol{F}_{\text{vdW}}$, always acts along the line connecting two spheres and can be expressed in terms of the gradient of the van der Waals potential. \citet{hamaker1937london} derived an expression for the van der Waals potential using the principle of pairwise additivity; however, his calculation neglects electromagnetic retardation, which becomes significant when the separation distance between particles is of the order of, or exceeds, the London wavelength $\lambda_L (\approx 0.1$ \textmu m$)$. In this work, we determine the van der Waals force from the expression for the retarded van der Waals potential given in \citet{zinchenko1994gravity}. The retarded form of $\boldsymbol{F}_{\text{vdW}}$ depends on $r/a^*$, $\kappa$, $A_H$ (the Hamaker constant), and $N_L = 2 \pi \left(a_1+a_2\right)/\lambda_L = 2 \pi a_1 \left(1+\kappa\right)/\lambda_L$ (quantifies the dimensionless particle separation distance beyond which retardation effects become important). The non-dimensional van der Waals force $\boldsymbol{f}_v$, which we will use in subsequent analysis, is defined as $\boldsymbol{f}_v=\boldsymbol{F}_{\text{vdW}}/(A_H/a^*)=-f_v \boldsymbol{\hat{e}}_r$.

In line with the work of \citet{patra2023collision}, we account for the finite dielectric constants of the particles in our collision calculations. We assume that the particles are perfect dielectrics carrying prescribed free charges uniformly distributed over their surfaces, with no free charge within their interiors. The prescribed free surface charge is assumed not to redistribute during a particle encounter, and the net charge carried by each particle therefore remains constant. This assumption does not, however, imply a spatially uniform electrostatic response. The electric field generated by one particle polarizes the other, giving rise to a spatially non-uniform bound surface charge that varies with the instantaneous particle configuration. Thus, while the prescribed free surface charge remains fixed and uniform, the total surface charge distribution is generally non-uniform and evolves as the particles approach one another.

For particles suspended in air, the assumption that the net particle charge remains effectively constant during a collision event can be justified by comparing the characteristic encounter time with the time scale for charge relaxation through the surrounding gas. In the present uniaxial compressional flow, the characteristic particle relative velocity is $V_{\textrm{rel}}\sim\dot{\gamma}a^*$. Consequently, the corresponding collision time scale is given by  $t_{\textrm{coll}}=a^*/V_{\textrm{rel}}=1/\dot{\gamma}$; for $\dot{\gamma}=25\,\mathrm{s^{-1}}$, this gives $t_{\textrm{coll}} \approx 0.04\,\mathrm{s}$. In a weakly conducting gas, charge relaxation occurs on the Maxwell relaxation time $\tau=\epsilon_0/\sigma$, where $\epsilon_0 = 8.85 \times 10^{-12}$ C$^2$N$^{-1}$m$^{-2}$ is the permittivity of free space and $\sigma$ is the electrical conductivity of the surrounding medium. For clear air at sea level, $\sigma \approx 2.3\times10^{-14}\,\mathrm{S\,m^{-1}}$, corresponding to $\tau\approx6.5\,\mathrm{min}$. The lower electrical conductivity within warm clouds increases this relaxation time to approximately $20$ min to $2$ h \citep{pruppacher1997microphysics}. Therefore $t_{\textrm{coll}} \ll \tau$, implying that charge leakage through the surrounding gas is negligible over the collision time scale. This separation of time scales supports the assumption that the prescribed net charges remain constant during the collision process.  For an aqueous cloud droplet, finite internal conductivity may facilitate the redistribution of free charge along the interfacial region.  However, this effect is not accounted for in the current perfect-dielectric formulation and would require the coupling of ionic charge transport to the evolving electrostatic field.

Within this framework, we employ the formulation of \citet{khachatourian2014electrostatic}, who calculated the electrostatic interaction force between two charged dielectric spheres by deriving the potential field in a bispherical coordinate system. The non-dimensional electrostatic force $\boldsymbol{f}_e$, defined as $\boldsymbol{f}_e = \boldsymbol{F}_e/[q_1^2/(4\pi\epsilon_0a_1^2)] = f_e \boldsymbol{\hat{e}}_r$, depends on the size ratio $\kappa$, charge ratio $\beta$, dielectric constants of sphere 1, sphere 2, and the surrounding fluid medium, denoted as $k_1$, $k_2$, and $k_m$, respectively. For completeness, we briefly outline below the procedure used to compute the electrostatic force; further details are given in \citet{khachatourian2014electrostatic}.

The expression for the electrostatic force is given by:
\begin{eqnarray}
\boldsymbol{F}_e = -\hat{e}_r \ 4\pi\epsilon_0 \sum_{n=0}^{\infty}\mathcal{F}_n\Bigg[\frac{n}{2}\varPhi_{1,n-1}e^{\eta_1}-\left(n+\frac{1}{2}\right)\varPhi_{1,n}+\left(\frac{n+1}{2}\right)\varPhi_{1,n+1}e^{-\eta_1}\Bigg]\varPhi_{2,n}.
\end{eqnarray}
The potential coefficients $\varPhi_{1,n}$ and $\varPhi_{2,n}$, which account for the mutual dielectric polarization of the two spheres, are obtained from two coupled recursive relations. One of them is
\begin{eqnarray}
&&\Bigg[\left(n+\frac{1}{2}\right)\varPhi_{1,n}\cosh\eta_1-\frac{n}{2}\varPhi_{1,n-1}-\frac{n+1}{2}\varPhi_{1,n+1}\Bigg]\left(k_m+k_1\right)+ \frac{\sinh\eta_1}{2}\left(k_m-k_1\right)\varPhi_{1,n} \nonumber \\
&&-\Bigg[\left(n+\frac{1}{2}\right)\varPhi_{2,n}\mathcal{F}_n\cosh\eta_1-\frac{n}{2}\varPhi_{2,n-1}\mathcal{F}_{n-1}-\frac{n+1}{2}\varPhi_{2,n+1}\mathcal{F}_{n+1}\Bigg]\left(k_m-k_1\right) \nonumber \\
&&+\mathcal{F}_n\frac{\sinh\eta_1}{2}\left(k_m-k_1\right)\varPhi_{2,n} = \frac{\sqrt{2}\mathcal{A} q_1 e^{-\left(n+\frac{1}{2}\right)\eta_1}}{4\pi\epsilon_0 a_1^2},
\label{Coefficient_of_potential_expression}
\end{eqnarray}
with the other one obtained by interchanging subscripts 1 and 2 in (\ref{Coefficient_of_potential_expression}). The expressions for $\eta_1$, $\eta_2$, $\mathcal{A}$ and $\mathcal{F}_n$ involved in the above two recursive relations are given by
\begin{eqnarray}
&&\eta_1 = \cosh^{-1}\Bigg[\frac{\left(1+\kappa\right)(\xi+2)}{4}+\frac{\left(1-\kappa^2\right)}{\left(1+\kappa\right)(\xi+2)}\Bigg],\\
&&\eta_2 = \cosh^{-1}\Bigg[\frac{\left(1+\kappa\right)(\xi+2)}{4\kappa}-\frac{\left(1-\kappa^2\right)}{\kappa\left(1+\kappa\right)(\xi+2)}\Bigg],\\
&&\mathcal{A} = a_1\sqrt{\frac{\left(1+\kappa\right)^2(\xi+2)^2}{16} + \frac{\left(1-\kappa^2\right)^2}{\left(1+\kappa\right)^2(\xi+2)^2}-\frac{\left(1+\kappa^2\right)}{2}},\\
&&\mathcal{F}_n = e^{-\left(n+\frac{1}{2}\right)\left(\eta_1+\eta_2\right)}.
\end{eqnarray}
Solving the two coupled recursive relations for $\varPhi_{1,n}$ and $\varPhi_{2,n}$ and substituting the resulting coefficients into the force expression above yields the electrostatic interaction force at each particle separation.

To simplify the analysis, we express the particle relative velocity equation in a spherical coordinate system, with the origin located at the center of the test droplet. We nondimensionalize the relative velocity equation using characteristic scales for length, velocity, and time, which are defined as $a^*$, $\dot{\gamma}a^*$, and $\dot{\gamma}^{-1}$, respectively. Because of this scaling, the nondimensional radial separation between the centers of the two spheres can vary from $2$ to $\infty$, and henceforth we denote this nondimensional radial distance by $r$. The sphere with a dimensionless radius of $r = 2$ is referred to as the collision sphere. In contrast, at $r = \infty$, the influence of one sphere on the other becomes negligible. We assume $a_1 > a_2$, allowing the size ratio to vary within the range $(0,1]$. Performing the vector and tensor operations in Eq.~(\ref{Eq: relative_velocity}) yields the components of the dimensionless relative velocity defined by $\boldsymbol{v}=\boldsymbol{V}_{12}/\dot{\gamma}a^*$, expressed in the radial ($r$), polar ($\theta$), and azimuthal ($\phi$) directions as follows:
\begin{eqnarray}
    v_r &=& -r\left(1-A\right)\left(3\cos^{2}\theta-1\right) + N_e G f_e - N_v G f_v, \label{Eq:radial_comp} \\
    v_{\theta} &=& 3r\left(1-B\right)\sin\theta \cos\theta, \label{Eq:theta_comp}\\
    v_{\phi} &=& 0\label{Eq:phi_comp}. 
\end{eqnarray}
Here, $N_e$ and $N_v$ are the dimensionless quantities that measure the strengths of the electrostatic and retarded van der Waals forces, respectively, relative to the background compressional flow: 
\begin{eqnarray}
    N_e &=& \frac{q_1^2}{12\pi^2\mu_f\epsilon_0 a_1^4\dot{\gamma}\kappa}, \qquad \textrm{and} 
    \label{Eq:Ne} \\
    N_v &=& \frac{2 A_H}{3\pi\dot{\gamma}\mu_f\kappa(1+\kappa)a_1^3}.
    \label{Eq:Nv}
\end{eqnarray}
To place these parameters in an atmospheric context, we estimate $\dot{\gamma}$ from the turbulent dissipation rate $\varepsilon$ using the relation introduced in \S \ref{sec:Introduction}. As a result, the computed values of $N_e$ and $N_v$ correspond to strain rates representative of atmospheric turbulence, while the underlying calculations adhere to the assumptions of a steady, uniaxial compressional flow.

For warm clouds, reported dissipation rates are $\varepsilon \sim 10^{-3}$--$10^{-1}\,\mathrm{m^2\,s^{-3}}$ \citep{grabowski2013growth}, corresponding to $\dot{\gamma} \sim 2$--$20\,\mathrm{s^{-1}}$ for $\nu_f \approx 1.8\times10^{-5}\,\mathrm{m^2\,s^{-1}}$. Cloud droplets with radii $a_1 \sim 10$--$20\,\upmu\mathrm{m}$ may carry charges of order $10$--$10^2\,e$ \citep{pruppacher1997microphysics}. Here, $e=1.602 \times 10^{-19}$ C is the elementary charge. For these ranges, Eq. (\ref{Eq:Ne}) gives $N_e=O(10^{-5})$--$O(10^0)$. Electrostatic interactions may therefore become appreciable for relatively highly charged small droplets subjected to weak local strain. Similarly, using  $A_H \sim 4\times10^{-20}\,\mathrm{J}$ for water-based particles  \citep{israelachvili2011intermolecular}, Eq. (\ref{Eq:Nv}) gives $N_v = O(10^{-3})$--$O(10^{-1})$.

Volcanic plumes represent a markedly different energetic regime. Proximal volcanic plumes may exhibit dissipation rates of $\varepsilon \sim 10^{-1}$--$10^2\,\mathrm{m^2\,s^{-3}}$ \citep{textor2004comment}, corresponding to $\dot{\gamma} \sim 20$--$700\,\mathrm{s^{-1}}$. Measurements of silicate ash particles with radii $a_1 \sim 30$--$100\,\upmu\mathrm{m}$ indicate charges of approximately $q_1 \sim 10^{-15}$--$10^{-13}\,\mathrm{C}$  \citep{harrison2010self,mendez2016effects}. For these representative ranges, Eq. (\ref{Eq:Ne}) gives $N_e = O(10^{-3})$--$O(10^4)$. Similarly, using $A_H \sim (3$--$10) \times 10^{-20}\,\mathrm{J}$ for silicate particles, Eq.(\ref{Eq:Nv}) gives $N_v = O(10^{-7})$--$O(10^{-3})$. Particle inertia becomes significant across much of the volcanic ash parameter range. For silicate ash with $\rho_p\approx2.5\times10^3\,\mathrm{kg\,m^{-3}}$, the Stokes number varies roughly between $O(1)$ and $O(10^2)$ for the above particle sizes and strain rates. Consequently, the current inertialess formulation applies solely to sufficiently fine ash particles. For example, $St\lesssim0.3$ requires $a_1\lesssim20\,\upmu\mathrm{m}$ at $\dot{\gamma}\approx20\,\mathrm{s^{-1}}$ and $a_1\lesssim4\,\upmu\mathrm{m}$ at $\dot{\gamma}\approx700\,\mathrm{s^{-1}}$. Therefore, the estimates of $N_e$ and $N_v$ for coarser ash particles should be interpreted as indicative of the interaction strengths achievable in volcanic plumes, rather than as parameter ranges to which the current inertialess pair dynamics can be quantitatively applied.

These estimates demonstrate that electrostatic and van der Waals interactions depend strongly on particle size, charge, and local strain rate in both scenarios. They also highlight a fundamental contrast: while electrostatic interactions typically remain weak in warm clouds, they may become comparable to or exceed the local flow-induced contribution in electrically active volcanic environments.

It is important to note that temperature is not varied independently in the present formulation. While calculating the above non-dimensional parameters, the influence of temperature enters primarily through the temperature-dependent density, viscosity, and molecular mean free path of the suspending medium, as well as the particle/droplet density. Similarly, the charging process itself is not modeled dynamically; instead, $q_1$ and $q_2$ are prescribed parameters, with representative charge ranges chosen separately for cloud droplets and volcanic ash particles. Thus, the present calculations quantify the collision dynamics for specified particle charges rather than the temperature-dependent generation of those charges.

\subsection{Expressions for the collision rate and efficiency}\label{Expressions_for_the_collision_rate_and_efficiency}

The collision rate, denoted by $K_{12}$, quantitatively measures the rate at which particles of radii $a_1$ and $a_2$ with their respective number densities $n_1$ and $n_2$ collide with each other per unit volume. Mathematically, it can be expressed as the flux of particles into the collision surface, as given by the following equation: 
\begin{equation}
    K_{12} = -\frac{1}{8}n_1n_2 \dot{\gamma} \left(a_1+a_2\right)^3 \int_{(r=2)\&\left(\boldsymbol{v}\mathbf{\cdot}\boldsymbol{n}<0\right)}  \left(\boldsymbol{v}\mathbf{\cdot}\boldsymbol{n}\right) P(r) dA,
\label{Collision_rate_general_dimensional}
\end{equation}
where $P(r)$ denotes the pair distribution function and $\boldsymbol{n}$ is the outward unit normal vector at the collision surface. The condition $\boldsymbol{v}\mathbf{\cdot}\boldsymbol{n}<0$ ensures that only inward-directed radial relative velocities at the collision surface contribute to the collision rate. For a dilute dispersion, the pair-distribution function is governed by the quasi-steady Fokker-Planck equation applicable to the spatial region external to the contact surface:
\begin{equation}
    \boldsymbol{\nabla}\mathbf{\cdot}\left(P\boldsymbol{v}\right) = 0.
\label{Probability_conservation}
\end{equation}
As the particle motions in the far field become uncorrelated, this leads to the boundary condition: $P \rightarrow 1$ as $r \rightarrow \infty$. For calculation purposes, we consider $r=r_{\infty}$, representing a large, yet finite, separation distance.

The two equations above do not include any diffusive flux, permitting the evaluation of the collision rate through trajectory analysis. By applying Eq.~(\ref{Probability_conservation}) in conjunction with the divergence theorem, the integral in Eq.~(\ref{Collision_rate_general_dimensional}) can be reformulated over the closed surface that bounds all trajectories originating at $r=r_{\infty}$ and terminating at $r=2$. As a result, the flux through the cross-section of this volume at $r=r_{\infty}$ - referred to as the upstream interception area, $A_c$ - defines the collision rate. Given that the pair-distribution function satisfies the condition $P=1$ at $r=r_{\infty}$, Eq.~(\ref{Collision_rate_general_dimensional}) reduces to
\begin{equation}
   K_{12} = -\frac{1}{8}n_1n_2 \dot{\gamma} \left(a_1+a_2\right)^3\int_{A_c}\left(\boldsymbol{v}\mathbf{\cdot}\boldsymbol{n}'\right)|_{r_{\infty}} dA,
\label{Collision_rate_upstream_interception_area} 
\end{equation}
where $\boldsymbol{n}'$ denotes the outward unit normal vector on the area elements of $A_c$. It is important to note that the term $\left(\boldsymbol{v}\mathbf{\cdot}\boldsymbol{n}'\right)|_{r_{\infty}}$ can be evaluated by considering the contribution from the background flow only, as electrostatic and van der Waals forces are negligible at $r=r_{\infty}$.

The collision rate calculated without considering any interactions, whether hydrodynamic or non-hydrodynamic, is referred to as the ideal collision rate, denoted by $K_{12}^{0}$. For a uniaxial compressional flow, \citet{zeichner1977use} derived the following expression for the ideal collision rate:
\begin{equation}
    K^{0}_{12} = \frac{8\pi}{3\sqrt{3}}n_1n_2\dot{\gamma}\left(a_{1}+a_{2}\right)^{3}.
\label{ideal_collision_rate}
\end{equation}
Finally, the ratio of actual collision rate, $K_{12}$, to the ideal collision rate, $K_{12}^{0}$, defines the collision efficiency, denoted as $E_{12}$:
\begin{equation}
    E_{12} = \frac{K_{12}}{K_{12}^{0}}.
\label{Collision_efficiency}
\end{equation}

\section{Numerical method} \label{Numerical_method}

To calculate the collision rate, it is essential to determine the upstream interception area, as evident from Eq. (\ref{Collision_rate_upstream_interception_area}). We achieve this by employing a trajectory analysis methodology, which involves initializing a test sphere at the origin and evolving the satellite spheres to identify those that collide. Since the particle relative velocity does not depend on the azimuthal coordinate $\phi$, we can effectively describe the dynamics by analysing the relative trajectories in a generic $r\sin\theta-r\cos\theta$ plane. We find the relative trajectories of a particle pair by integrating the following dimensionless trajectory equation:
\begin{equation}
   \frac{d\theta}{dr} = \frac{v_{\theta}}{r v_r} = \dfrac{3\left(1-B\right)\sin\theta \cos\theta}{-r\left(1-A\right)\left(3\cos^{2}\theta-1\right) + N_e G f_e - N_v G f_v}.
\label{final_trajectory_equation} 
\end{equation}
This equation captures the relative motion of two spherical particles, influenced by the combined effects of a uniaxial compressional flow, hydrodynamic interactions, electrostatic forces, and van der Waals forces.

Colliding trajectories are defined as the paths of satellite sphere centers originating far upstream and terminating at the collision surface. In the far-field, these colliding trajectories constitute the upstream interception area $A_c$. Direct computation of colliding trajectories by choosing initial conditions over a spherical shell of dimensionless radius $r_{\infty}$ is computationally expensive, as most trajectories originating from this spherical shell do not reach the collision sphere. To address this limitation, we exploit the quasi-steady nature of the relative trajectory equation and perform backward integration using a fourth-order Runge–Kutta method, with initial conditions specified on the collision surface. At exactly $r=2$, however, the radial relative velocity vanishes because the relevant hydrodynamic mobilities $A=1$ and $G=0$ at the contact surface. To circumvent this singularity, we choose our initial conditions on a slightly expanded sphere of radius $2+\delta$, where $\delta$ is a slight offset from the collision surface. We obtain converged results with reasonable computational cost for $\delta = 10^{-6}$. Furthermore, as evident from Eq.~(\ref{Collision_rate_general_dimensional}), only trajectories with inward radial velocity at the collision sphere contribute to collisions; therefore, we further reduce computational effort by choosing initial conditions on the collision sphere where $v_r < 0$. 

Among these computed colliding trajectories, the limiting colliding trajectories exhibit maximum horizontal shifts from the compressional axis in the far field. The upstream interception areas for this problem are two identical circles whose radius is the far-field horizontal shift of these limiting colliding trajectories. After identifying the upstream interception areas, we calculate the collision efficiency. We use this numerical methodology to compute the collision efficiency for a range of values for $\kappa$, $Kn$, $\beta$, $N_e$, and $N_v$.

\section{Results and discussion} \label{sec: trajectory calculations}

\subsection{Electrostatic modification of charged-tracer collisions} \label{subsec:collision_rate_between_two_charged_tracers}

Before turning to the fully coupled hydrodynamic-electrostatic problem, we first examine how electrostatic interactions alone modify the collision rate between two charged tracers in a uniaxial compressional flow. We consider two tracers of unequal size and charge and neglect hydrodynamic interactions by taking $A, B \rightarrow 0$ and $G \rightarrow 1$. Retaining the monopole-monopole interaction and the leading monopole-induced-dipole correction, the radial component of the relative velocity is given by
\begin{equation}
    v_{r} = -r \left(3\cos^{2}\theta - 1\right) + \frac{4 N_{e}}{(1+\kappa)^{2}} \left[\frac{\beta}{r^{2}} - \frac{16\,(\beta^{2} + \kappa^{3})\,\mathscr{K}}{(1+\kappa)^{3}\,r^{5}} \right].
    \label{Eq:non_dim_charged_tracer}
\end{equation}
In the above equation, the parameter $\mathscr{K}=(k-1)/(k+2)$, where $k=k_1=k_2$, denotes the Clausius–Mossotti factor, which quantifies the relative polarizability of the particle in its surrounding medium; for water droplets in air, $\mathscr{K} \approx 0.96$.

To compute the critical angle $\theta_{c}$ associated with the limiting trajectory that reaches the collision surface, we set $v_{r} = 0$ at $r = 2$ in Eq. \eqref{Eq:non_dim_charged_tracer}. This gives
\begin{equation}
    \theta_{c} = \cos^{-1}\left[\frac{1}{\sqrt{3}}\left\{1 + \frac{\tilde{N}_{e}}{8} \left(1 - \frac{2(\beta^{2} + \kappa^{3})\,\mathscr{K}}{\beta\,(1+\kappa)^{3}} \right) \right\}^{1/2} \right],
    \label{Eq:critical_angle}
\end{equation}
where we have defined the rescaled electrostatic parameter as $\tilde{N}_{e} = 4 \beta N_e/(1+\kappa)^2$.

\begin{figure}
\centering
\includegraphics[width=1.0\textwidth]{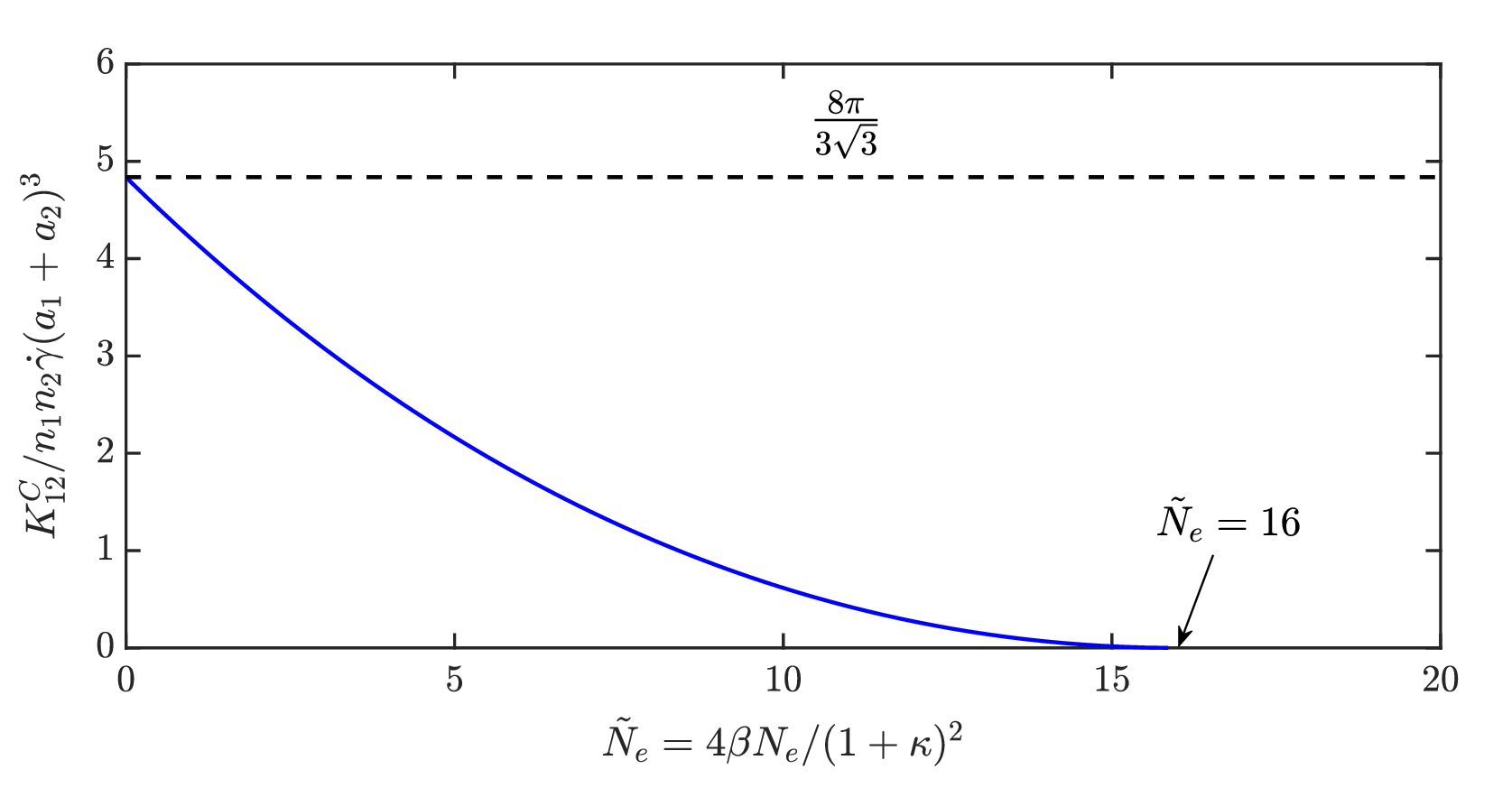}
\caption{Variation of the scaled collision rate, $K_{12}^{\mathrm{C}}/[n_1n_2\dot{\gamma}(a_1+a_2)^3]$, with the scaled electrostatic parameter $\tilde{N}_e$ for a pair of like-charged tracers interacting through Coulombic repulsion in a uniaxial compressional flow. The collision rate decreases monotonically with increasing $\tilde{N}_e$ and vanishes at the critical value $\tilde{N}_e=16$, beyond which Coulombic repulsion prevents particle contact. The horizontal dashed line represents the ideal collision rate for a uniaxial compressional flow alone, $K_{12}^{0}/[n_1n_2\dot{\gamma}(a_1+a_2)^3]=8\pi/(3\sqrt{3})$.}
\label{fig:ideal_collision_rate_charged_tracer}
\end{figure}

The purely Coulombic limit follows directly by setting $\mathscr{K}=0$, thereby retaining only monopole-monopole repulsion. In this limit, the relative velocity field is solenoidal, so the pair-conservation equation $\boldsymbol{\nabla}\boldsymbol{\cdot}(P\boldsymbol{v})=0$, together with $P \rightarrow 1$ in the far field, gives $P=1$ for all values of $r$. The collision rate can therefore be evaluated directly from the inward flux across the collision sphere, yielding
\begin{equation}
    K_{12}^{\mathrm{C}} = \frac{\pi}{2} \left[\frac{16}{3\sqrt{3}} \left(1+\frac{\tilde{N}_{e}}{8}\right)^{3/2} -\tilde{N}_{e} \right] n_1 n_2 \dot{\gamma}(a_1+a_2)^3.
    \label{Eq:Coulombic_collision_rate}
\end{equation}
Here, $K_{12}^{\mathrm{C}}$ denotes the collision rate in the presence of Coulombic repulsion and is distinct from the ideal collision rate $K_{12}^{0}$ defined in \S \ref{Expressions_for_the_collision_rate_and_efficiency}, for which all kinds of interactions are absent. Indeed, for $\tilde{N}_{e} = 0$, $K_{12}^{\mathrm{C}} = K_{12}^{0}$, recovering the classical ideal collision rate for a uniaxial compressional flow. Coulombic repulsion therefore monotonically reduces the collision rate from its ideal value. At $\tilde{N}_{e}=16$, the region $v_r < 0$ on the collision sphere collapses to the compressional axis, while for $\tilde{N}_{e} > 16$ no trajectory originating in the far field reaches contact and hence $K_{12}^{\mathrm{C}}=0$ (see Figure \ref{fig:ideal_collision_rate_charged_tracer}).

\begin{figure}
\centering
\includegraphics[width=1.0\textwidth]{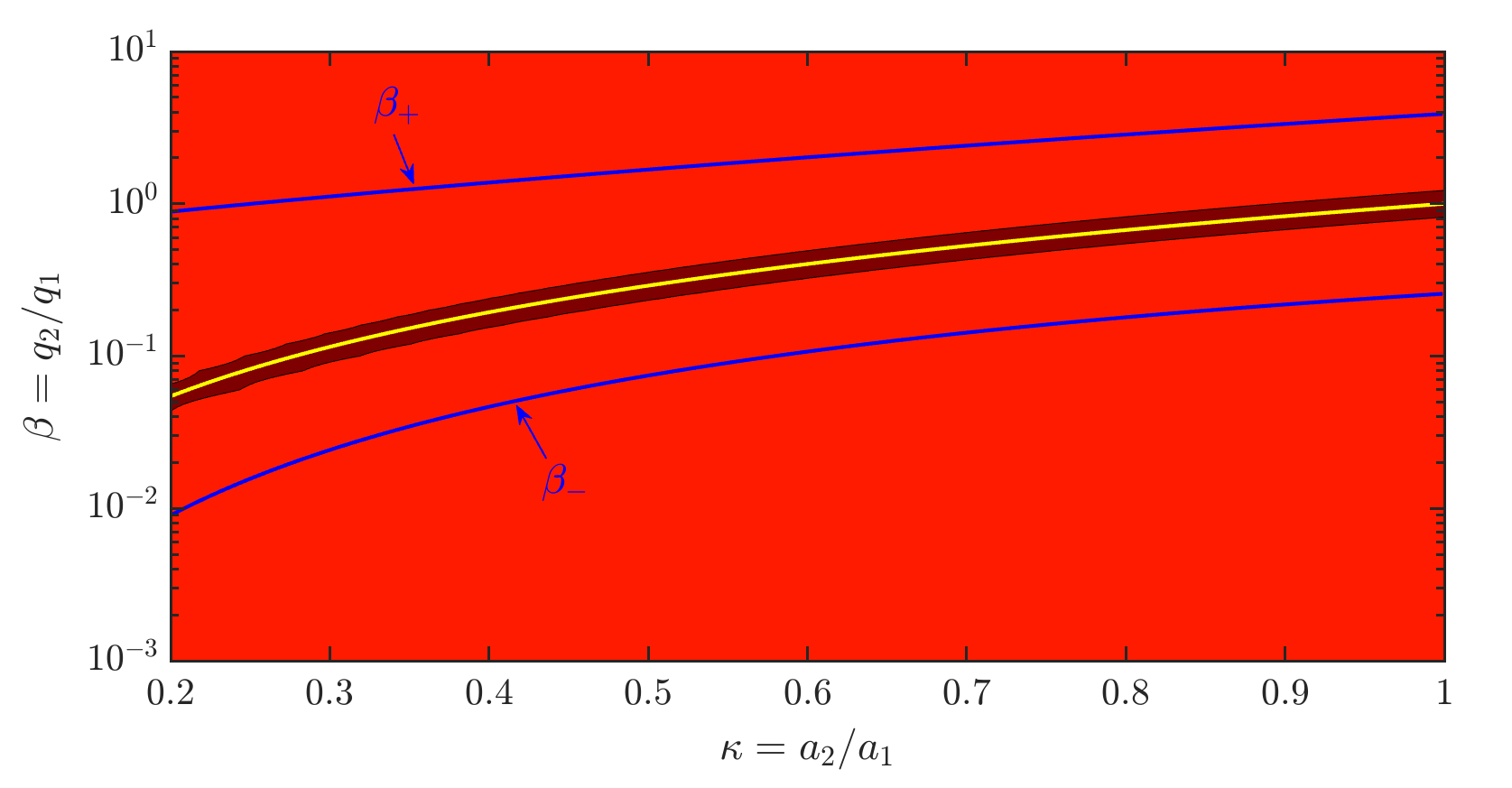}
\caption{Electrostatic interaction force near contact ($\xi=10^{-6}$) in the $(\kappa,\beta)$ parameter space for two like-charged water droplets in air. The two curves $\beta_{-}$ and $\beta_{+}$ represent the attraction-repulsion boundaries predicted by the monopole-dipole approximation, with repulsion occurring for $\beta_{-} < \beta <\beta_{+}$. The background map shows the sign of the full electrostatic force calculated using the finite-dielectric formulation of \citet{khachatourian2014electrostatic}; the dark-red region denotes repulsive interactions, whereas interactions are attractive outside this region. The yellow curve represents the condition for which the electrostatic interaction between two like-charged conducting spheres is repulsive.}
\label{fig:force_at_contact}
\end{figure}

Polarization effects can significantly alter the collision rate due to Coulombic repulsion. At contact, the leading electrostatic contribution in (\ref{Eq:non_dim_charged_tracer}) changes sign when
\begin{equation}
    2(\beta^2+\kappa^3)\mathscr{K} = \beta(1+\kappa)^3, \label{Eq:monopole_dipole_boundary}
\end{equation}
which leads to the two distinct boundaries defined by
\begin{equation}
    \beta_{\pm} = \frac{(1+\kappa)^3 \pm \sqrt{(1+\kappa)^6-16\mathscr{K}^2\kappa^3}} {4\mathscr{K}}.
    \label{Eq:monopole_dipole_roots}
\end{equation}
The electrostatic force map shown in Figure \ref{fig:force_at_contact} illustrates how the sign of the near-contact interaction depends on the particle size and charge ratios. Within the range where $\beta_- < \beta < \beta_+$, the interaction is repulsive, whereas outside this interval, the polarization-induced attraction dominates. These two curves are superimposed on the full electrostatic force at contact, obtained from the finite-dielectric formulation of \citet{khachatourian2014electrostatic}. The complete calculations reveal a substantially narrower repulsive region compared to the monopole-dipole approximation, demonstrating that higher-order polarization contributions strengthen the near-field attraction and significantly modify the boundary between attraction and repulsion. Additionally, in Figure \ref{fig:force_at_contact}, the yellow line represents the condition under which the interaction between two like-charged conducting spheres becomes repulsive, providing a point of reference for comparison.

In contrast to the purely Coulombic interaction, the polarization contribution renders the relative velocity field non-solenoidal, so that $P \neq 1$ at finite separation. Consequently, the collision rate cannot be accurately estimated based solely on the electrostatic force at contact or the critical angle. Furthermore, the sign of the electrostatic force at contact does not exclusively dictate whether a pair of particles will collide; it merely indicates whether the electrostatic interactions locally facilitate or impede their radial approach at that point. In the complete problem, the electrostatic force further competes with the imposed compressional flow and lubrication resistance. We therefore employ the full finite-dielectric electrostatic interaction in the calculations that follow and determine the collision efficiency by analysing pair trajectories rather than from the truncated far-field interaction.

\subsection{Critical electrostatic strength for collision suppression}
\label{subsec:Critical_Ne}

The preceding analysis of charged tracers demonstrated that sufficiently strong Coulombic repulsion can entirely inhibit collisions in a uniaxial compressional flow. We now examine whether a similar threshold exists when considering hydrodynamic interactions and the complete electrostatic interaction between finite-size dielectric spheres. We denote this threshold by $(N_e)_c$, defined as the critical value of $N_e$ above which no trajectory starting from the far field can reach the collision surface. Given that the compressional axis (i.e., $\theta=0$) provides the maximum inward flow-induced radial velocity, the absence of inward motion along this axis determines the limiting condition for collision. For prescribed values of $\kappa$, $\beta$, and $Kn$, we compute $(N_e)_c$ by minimizing $2r(1-A)/(G f_e)$ over all separations for which $f_e > 0$. Mathematically, this is expressed as:
\begin{equation}
    (N_e)_c = \min_{\substack{r\geq 2\\f_e(r)>0}} \left[ \frac{2r(1-A)}{Gf_e} \right]. \label{Eq:critical_Ne_minimum}
\end{equation}
At the critical separation $r_c$, the radial relative velocity satisfies $v_r(r_c,\theta=0)=0$ and  $\left. \partial v_r/\partial r \right|_{r=r_c}=0$. It is important to note that the minimization is confined to those separations where the electrostatic force is repulsive, as an attractive electrostatic force reinforces rather than opposes the inward motion. Physically, $(N_e)_c$ measures the strength at which the largest electrostatic barrier encountered during approach first balances the inward motion induced by the compressional flow. For $N_e < (N_e)_c$, the flow can overcome this barrier, thereby permitting the occurrence of collisions. Conversely, for $N_e>(N_e)_c$, a region with $v_r>0$ develops during approach, effectively preventing particles arriving from the far field from reaching the collision sphere.

\begin{figure}
    \centering
    \includegraphics[width=1.0\textwidth]{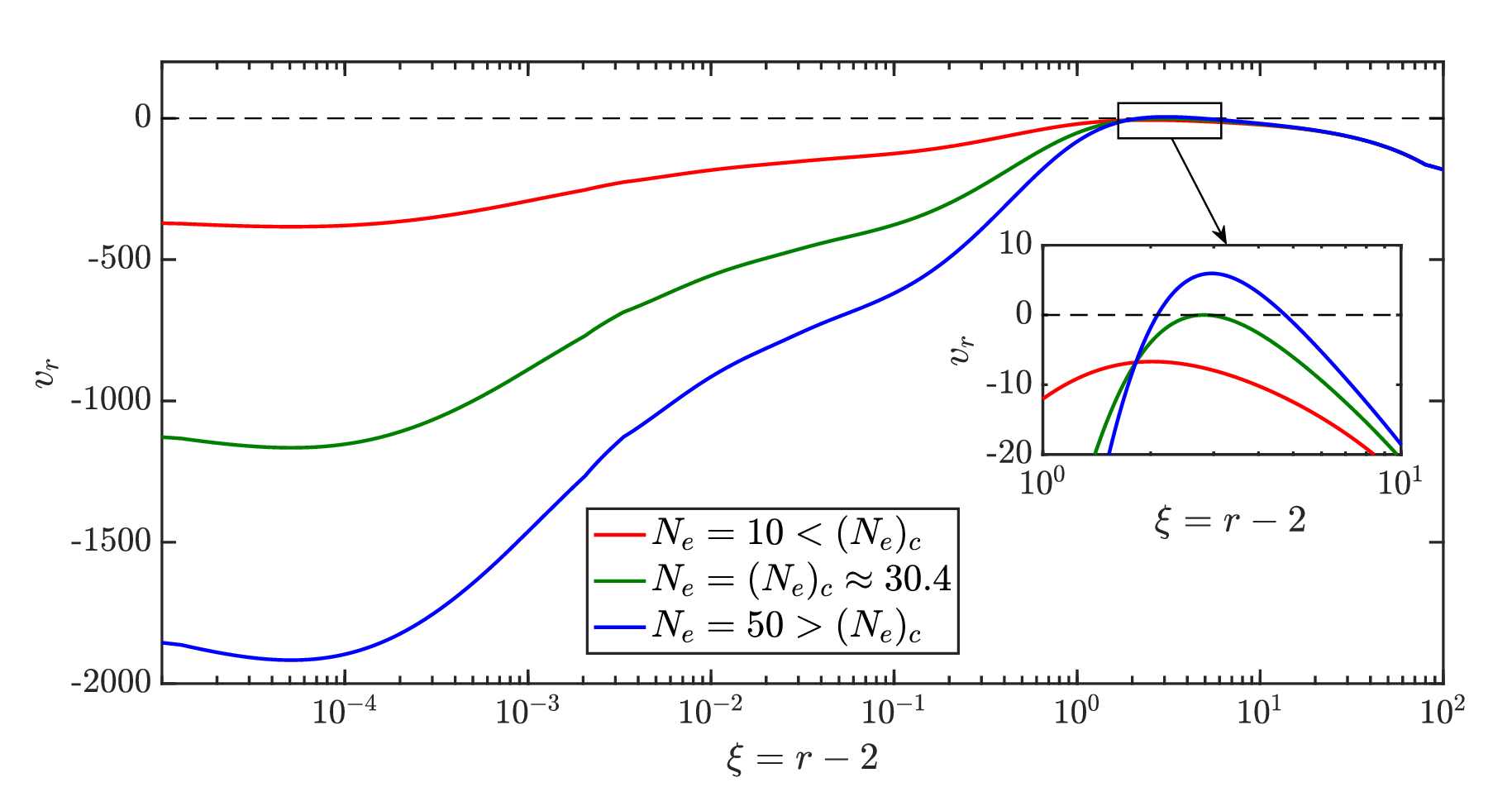}
    \caption{The radial component of the relative velocity, $v_r$, along the compressional axis ($\theta=0$) as a function of the dimensionless surface-to-surface separation $\xi$ for $N_e$ less than, equal to and greater than $(N_e)_c$, with $Kn=10^{-2}$, $\kappa=0.5$, $\beta=10$, $k_1=k_2=80$ and $N_v=0$. The inset magnifies the region near $v_r=0$: at $N_e=(N_e)_c$, $v_r$ reaches zero at a single separation, whereas for $N_e>(N_e)_c$ it becomes positive over a finite interval of $\xi$. This positive $v_r$ region represents an electrostatic repulsive barrier.}
    \label{fig:vr_vs_xi}
\end{figure}

\begin{figure}
    \centering
    \includegraphics[width=1.0\textwidth]{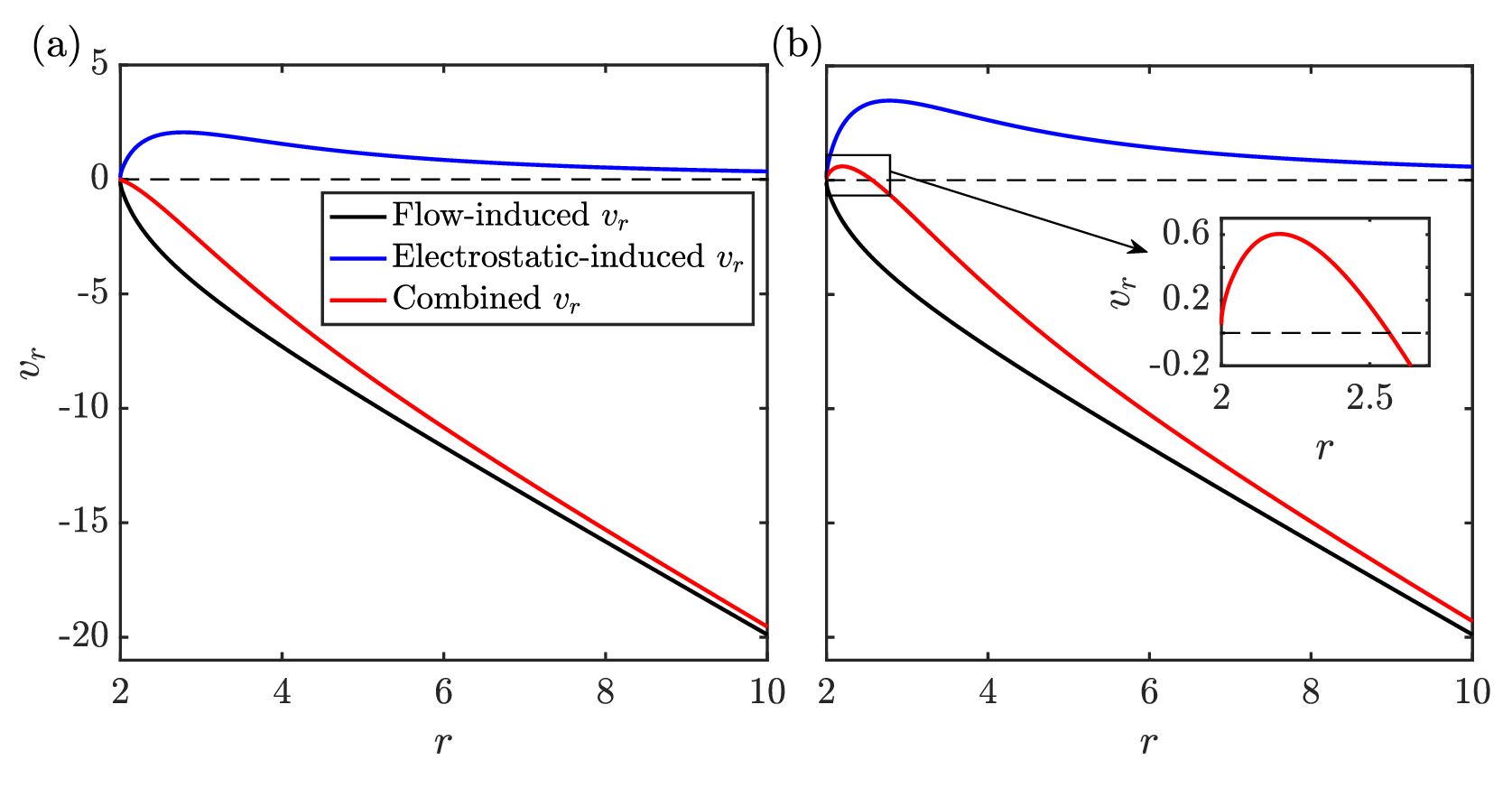}
    \caption{Radial component of the relative velocity, $v_r$, along the compressional axis ($\theta=0$) as a function of the dimensionless centre-to-centre distance $r$, for (a) $N_e<(N_e)_c$ and (b) $N_e>(N_e)_c$. The flow-induced (black), electrostatic (blue), and total (red) relative velocities are shown separately. These results are based on the parameters $Kn=10^{-2}$, $\kappa=0.6$, $\beta=0.4$, $k_1=k_2=80$ and $N_v=0$, for which $(N_e)_c \approx 70.1$.}
    \label{fig:repulsive_electrostatics_critical_Ne}
\end{figure}

We first consider combinations of $(\kappa,\beta)$ for which the electrostatic force is repulsive at large separations but becomes attractive in the near field. A representative example is $\kappa=0.5$ and $\beta=10$, for which we obtain $(N_e)_c \approx 30.4$ at $Kn=10^{-2}$ and $N_v=0$. Figure \ref{fig:vr_vs_xi} illustrates the corresponding radial relative velocity, $v_r$, as a function of the dimensionless surface-to-surface separation $\xi$ for values of $N_e$ that are below, at, and above $(N_e)_c$. Although the near-field electrostatic attraction promotes the final approach to contact, a particle pair approaching one another from the far field must first navigate through an intermediate region where the electrostatic interaction is repulsive. For $N_e<(N_e)_c$, the compressional flow is strong enough to overcome this repulsive barrier, resulting in $v_r$ remaining negative throughout the approach and allowing the particles to reach the near-field region where the electrostatic attraction further enhances the likelihood of collision. At $N_e=(N_e)_c$, the inward flow and electrostatic repulsion balance at the critical separation $\xi_c=r_c-2$, such that $v_r(\xi_c)=0$, while $v_r \leq 0$ elsewhere. As shown in the inset of Figure \ref{fig:vr_vs_xi}, increasing $N_e$ beyond $(N_e)_c$ results in the emergence of a finite interval over which $v_r>0$. This resulting outward relative motion prevents particles approaching from the far field from overcoming the repulsive electrostatic barrier, thus inhibiting their access to the attractive near-field region. Therefore, the presence of an attractive electrostatic force near contact does not automatically ensure a collision; rather, the likelihood of collision is determined by the interplay between the compressional flow and the separation-dependent electrostatic interaction throughout the particle approach.

\begin{figure}
  \centerline{\includegraphics[width=1.0\textwidth]{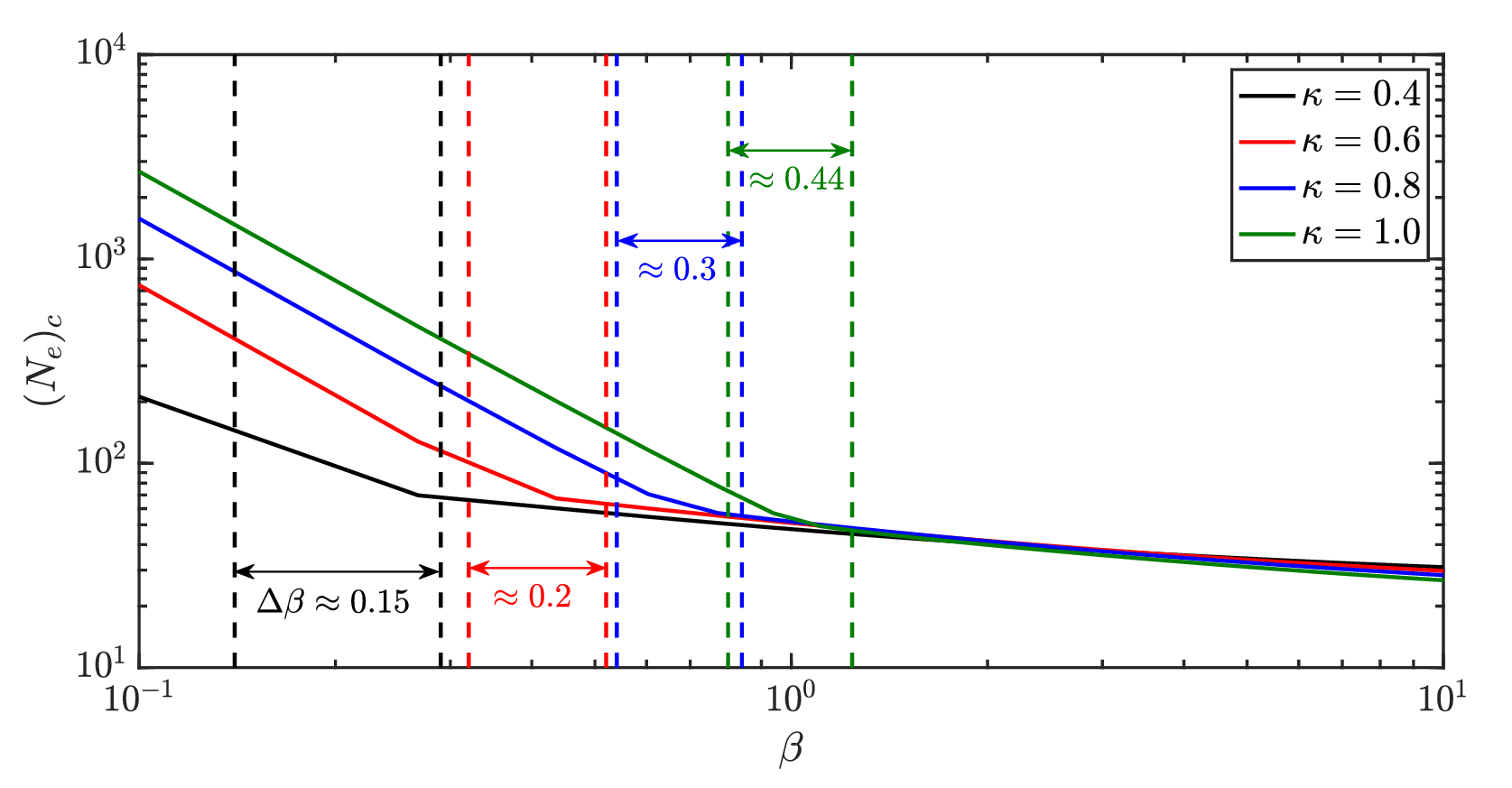}}
  \caption{Critical electrostatic parameter $(N_e)_c$ as a function of the charge ratio $\beta$ for different size ratios $\kappa$, with $Kn=10^{-2}$, $N_v=0$, and $k_1=k_2=80$. For each $\kappa$, the gap between the two vertical dashed lines of the same colour indicates the near-field repulsive interval $\Delta\beta$, determined from the complete electrostatic interaction between finite-size dielectric spheres. The changes in slope of the $(N_e)_c$ curves occur within these repulsive intervals. Although the repulsive intervals may appear narrower with increasing $\kappa$ because $\beta$ is plotted on a logarithmic scale, their actual width $\Delta\beta$ increases with $\kappa$.}
\label{fig:critical_Ne}
\end{figure}

A qualitatively distinct scenario emerges when the electrostatic interaction remains repulsive in the near field. For instance, at $\kappa=0.6$, the electrostatic force becomes repulsive in the near field for approximately $0.3 \lesssim \beta \lesssim 0.5$. Taking $\beta=0.4$ gives $(N_e)_c \approx 70.1$ for $Kn=10^{-2}$ and $N_v=0$.  Under these conditions, electrostatic repulsion prevents particles from approaching each other at all separation distances. When $N_e<(N_e)_c$, the inward velocity caused by the compressional flow is sufficiently large that the total radial velocity remains negative, allowing collisions to occur. Conversely, when $N_e>(N_e)_c$, the electrostatic repulsion forces surpass the contribution from the inward flow across a finite range of separations, resulting in $v_r > 0$ and inhibiting collisions. Figure \ref{fig:repulsive_electrostatics_critical_Ne} illustrates these two regimes.

The variation of $(N_e)_c$ with charge ratio is shown in Figure \ref{fig:critical_Ne} for four distinct values of $\kappa$. Across all examined cases, $(N_e)_c$ exhibits a decreasing trend with increasing values of $\beta$. Notably, the dependence of $(N_e)_c$ on $\beta$ undergoes a significant alteration within the regime where near-field interactions are repulsive in the $(\kappa,\beta)$ parameter space. In particular, the slope of the $(N_e)_c$ curves reveals marked changes within this repulsive band, with its width increasing as \(\kappa\) increases. These results show that the threshold for collision suppression is not set by the charge magnitude alone but depends jointly on the particle size and charge ratios.

\begin{figure}
      \centering
      \includegraphics[width=1.0\linewidth]{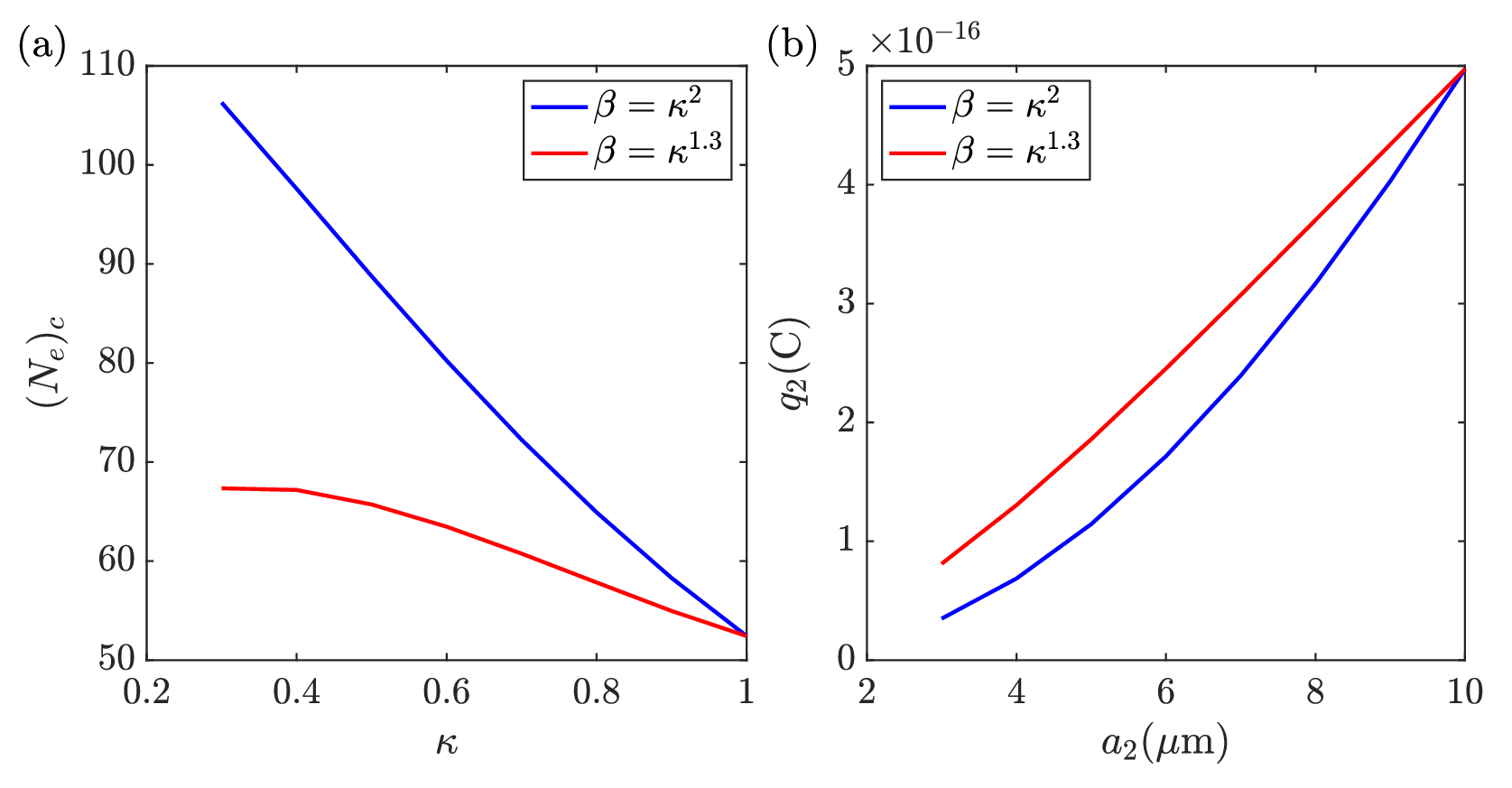}
      \caption{(a) Critical electrostatic parameter, $(N_e)_c$, as a function of the size ratio $\kappa$ for pairs of water droplets in air ($\lambda_0 = 0.1$ \textmu m), with the charge ratio-size ratio relationships $\beta=\kappa^2$ and $\beta=\kappa^{1.3}$ when $a_1=10$ \textmu m, $\dot{\gamma}=25\,\mathrm{s}^{-1}$ and $N_v=0$. (b) Corresponding critical charge on the test droplet as a function of its radius, above which electrostatic repulsion suppresses collision.}
      \label{fig:Ne_c_vs_kappa}
\end{figure}

These critical values of $N_e$ provide a useful basis for assessing the importance of electrostatic collision suppression under atmospheric conditions. To discuss this in the context of water droplets in warm clouds, we consider a satellite droplet of radius $a_1=10$ \textmu m and $\dot{\gamma}=25\,\mathrm{s}^{-1}$. We adopt two empirical relationships between droplet size and charge: $q_i \sim a_i^2$ and $q_i \sim a_i^{1.3}$ (where $i=1,2$), which lead to $\beta=\kappa^2$ and $\beta=\kappa^{1.3}$, respectively (see \citealt{pruppacher1997microphysics}). Utilizing these relationships, we determine $(N_e)_c$ over a range of size ratios $\kappa$. Figure \ref{fig:Ne_c_vs_kappa}(a) shows the resulting variation of $(N_e)_c$ with $\kappa$ for both $\beta-\kappa$ relationships. Figure \ref{fig:Ne_c_vs_kappa}(b) recasts these critical values of $N_e$ in terms of the charge on the test droplet, corresponding to the threshold for collision suppression. Based on our estimates, two equal-sized droplets, each with a radius of $10$ \textmu m, would require a charge of approximately $5 \times 10^{-16}\,\mathrm{C}$, or $3.1 \times 10^3e$, on each droplet for electrostatic repulsion to suppress their collision. This charge threshold is substantially larger than the representative average charge of a $10$ \textmu m cloud droplet (see \citealt{pruppacher1997microphysics}), suggesting that electrostatic repulsion is unlikely to suppress collisions between such weakly charged droplets under the conditions considered here. Volcanic ash particles exhibit a significantly broader range of electrostatic behaviors. For the representative particle sizes, charges and strain rates in volcanic plumes discussed in \S \ref{Trajectory_equations_for_a_particle_pair}, the resulting $N_e$ can approach or exceed $(N_e)_c$ for some like-charged particle pairs, although the present inertialess calculation applies quantitatively only to fine ash. Under sufficiently strong charging conditions, repulsive forces may inhibit particle collisions, while polarization-induced attractive forces can facilitate collisions in various regions of the size- and charge-ratio parameter space. These results highlight that $(N_e)_c$ should not be interpreted as a universal charge threshold; rather, it is contingent upon the complete electrostatic force profile and its competition with the imposed flow and hydrodynamic interactions.

\subsection{Trajectory topology and the critical threshold} \label{subsec:topology}

Having identified the critical electrostatic strength required to suppress collisions, we now proceed to investigate how increasing $N_e$ modifies the relative trajectories leading to this transition. Given that the relative velocity between the droplet pair remains unaffected by the azimuthal coordinate $\phi$, we show relative trajectories within a representative $r\sin\theta - r\cos\theta$ plane. To illustrate typical pair trajectories, we will use the parameter values $\kappa=0.5$, $\beta=10$, $Kn=10^{-2}$, $k_1=k_2=80$ and $N_v=0$. For these parameters, the analysis from the previous subsection gives $(N_e)_c \approx 30.4$.

\begin{figure}
    \centering
    \includegraphics[width=1.0\textwidth]{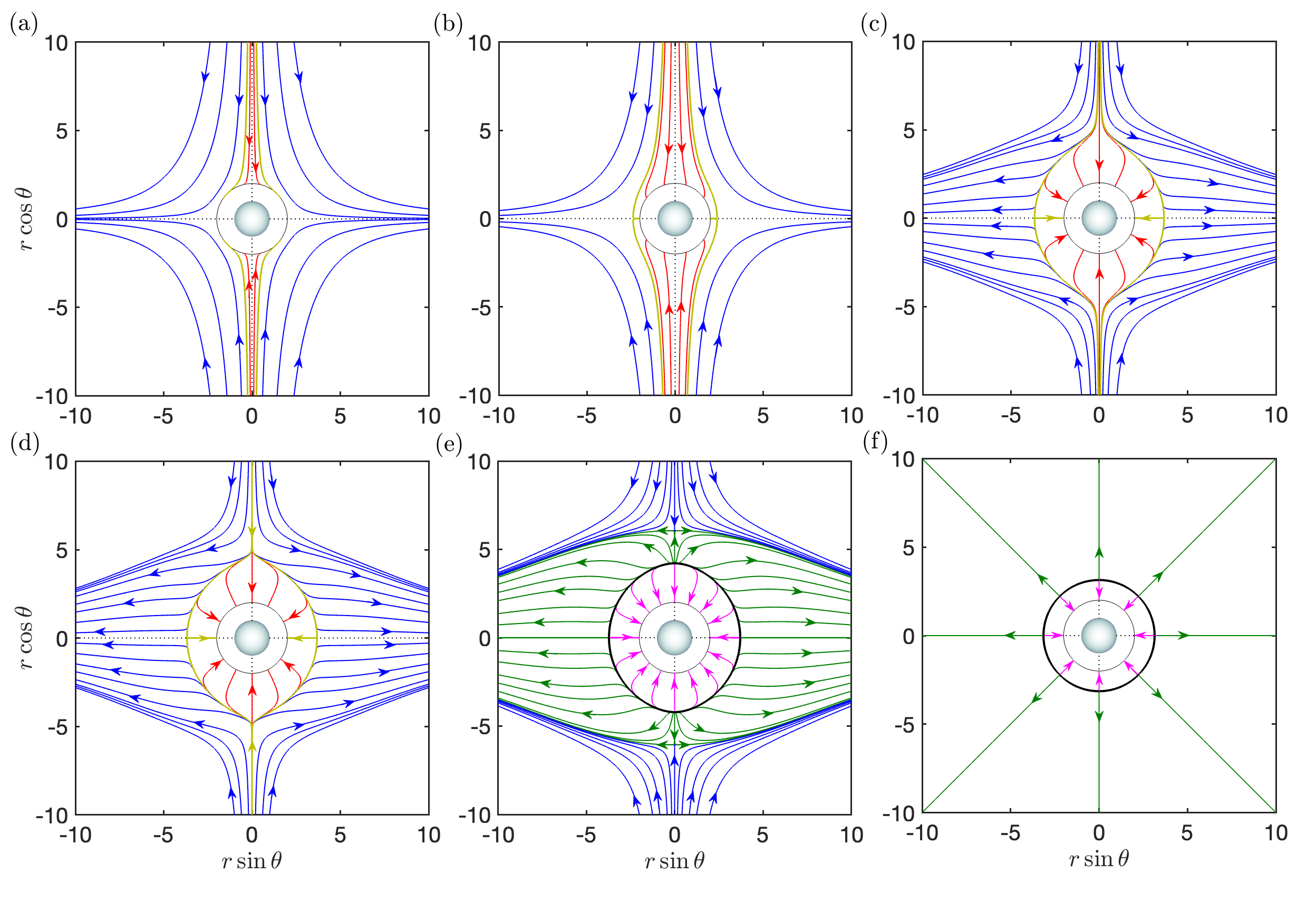}
    \caption{Relative trajectories of two like-charged spheres in a uniaxial compressional flow for $Kn=10^{-2}$, $\kappa=0.5$, $\beta=10$, $k_1=k_2=80$ and $N_v=0$ when (a) $N_e=0$, (b) $N_e=1$, (c) $N_e=25$, (d) $N_e=29$, (e) $N_e=40$ and (f) $N_e=\infty$. For these parameters, $(N_e)_c \approx 30.4$. The sphere at the centre is the test sphere, and the thin black circle denotes the projection of the collision sphere. Blue, red and golden lines denote open, colliding and limiting colliding trajectories, respectively. For $N_e>(N_e)_c$, the black lines delineate the isolated shell-like region created by the electrostatic barrier, the pink lines represent near-field trajectories driven inward by polarization-induced attraction, and the green lines denote trajectories outside this shell that diverge towards the far field. Colliding trajectories exist for $N_e<(N_e)_c$, as shown in (a-d), whereas no trajectory originating from the far field reaches the collision sphere for $N_e>(N_e)_c$, as shown in (e-f). In (f), the background flow is absent, and the relative motion is driven solely by electrostatic interactions.}
    \label{fig:trajectories_for_Ne}
\end{figure}

To establish a clear context, we first briefly review the pair trajectories when a uniaxial compressional flow alone ($N_e=0$) drives the relative motion between two hydrodynamically interacting spherical particles. Figure \ref{fig:trajectories_for_Ne}(a) shows the relative trajectories for the $N_e=0$ case. With non-continuum lubrication interactions, which permit particle contact in finite time, two distinct types of relative trajectories exist: open trajectories that approach from and return to the far field without reaching the collision sphere (blue lines), and colliding trajectories that approach from the far field and reach the collision sphere (red lines). These trajectories are fore-aft symmetric. Integrating Eq. \eqref{final_trajectory_equation} with $N_e=N_v=0$, gives
\begin{equation}
    \sin^2\theta(r)\cos\theta(r)=C\varphi(r),
    \label{relative_trajectory_expression_only_flow}
\end{equation}
where $C$ specifies an individual trajectory and
\begin{equation}
    \varphi(r)=\exp\left[\int_2^r
    \frac{3(B(r')-1)}{r'(1-A(r'))}\,\mathrm{d}r'\right].
    \label{varphi(r)_expression}
\end{equation}
Open and colliding trajectories correspond to $|C|>2/(3\sqrt{3})$ and $|C|<2/(3\sqrt{3})$, respectively, while the limiting colliding trajectories (golden lines) satisfy $|C|=2/(3\sqrt{3})$. These limiting trajectories act as separatrices between the open and colliding trajectories and intersect the collision sphere at $\theta=\theta_c$, $\pi-\theta_c$, $\pi+\theta_c$, and $2\pi-\theta_c$, where $\theta_c=\cos^{-1}(1/\sqrt{3})$. Accordingly, the limiting trajectory in the first quadrant can be obtained by initiating the backward integration from the collision sphere at $\theta=\theta_c$. These limiting colliding trajectories form the boundaries of the two upstream interception areas, which are circles whose centres lie on the compressional axis. The corresponding analytical expression for the collision efficiency is given by \citet{wang1994collision} as
\begin{equation}
    E_{12}=\exp\left[\int_2^\infty
    \frac{3(B(r)-A(r))}{r(1-A(r))}\,\mathrm{d}r\right].
    \label{Analytical_expression_collision_efficiency_only_flow}
\end{equation}

At finite $N_e$, electrostatic interactions modify the relative particle motion and, consequently, the topology of the pair trajectories and the locations at which the limiting colliding trajectories hit the collision sphere. Figure \ref{fig:trajectories_for_Ne}(b) shows the relative trajectories for $N_e=1$, which is well below $(N_e)_c$. Since the trajectories retain fore-aft symmetry at finite $N_e$, it is sufficient to analyse the relative trajectory equation numerically only over the interval $0 \leq \theta \leq \pi/2$. Unlike the $N_e=0$ case, the critical angle at which the limiting colliding trajectory hits the collision sphere is no longer prescribed by the purely hydrodynamic solution. We therefore determine $\theta_c$ numerically by scanning the interval $0 \leq \theta \leq \pi/2$ on the collision sphere and integrating \eqref{final_trajectory_equation} backward in $r$. For the parameters considered here, the limiting colliding trajectory originates at $\theta=\pi/2$. Consequently, colliding trajectories starting from the far field can hit the collision sphere at any point in the interval $0 \leq \theta \leq \pi/2$. Furthermore, the upstream interception areas for $N_e=1$ are larger than those for $N_e=0$, indicating an enhancement of the collision rate by electrostatic interactions for this case.

The influence of electrostatic repulsion becomes increasingly pronounced as $N_e$ approaches $(N_e)_c$. At $N_e=25$, the family of colliding trajectories forms a narrow neck region in the far field (see Figure \ref{fig:trajectories_for_Ne}(c)). Trajectories originating from this narrow neck region in the far field can reach the collision sphere upon forward integration by overcoming electrostatic repulsion at intermediate separations. An increase in $N_e$ to $29$ results in a further constriction of this neck region (see Figure \ref{fig:trajectories_for_Ne}(d)). This progressive narrowing of the neck region elucidates the mechanism for collision suppression within this trajectory-phase representation.

For $N_e>(N_e)_c$, the trajectory topology changes qualitatively. Figure \ref{fig:trajectories_for_Ne}(e) illustrates this regime for $N_e=40$. In this case, a pair approaching from the far field initially has $v_r<0$ because the inward radial motion induced by the compressional flow exceeds the opposing contribution from electrostatic repulsion. As the separation decreases, the repulsive electrostatic contribution becomes increasingly important until the two contributions balance at a finite separation, where $v_r=0$. A further decrease in separation results in $v_r>0$, which prevents the approaching pair from continuing up to the collision sphere. Correspondingly, backward integration from the collision sphere fails to yield trajectories that can reach the far field. Instead, the neck formation observed for $N_e<(N_e)_c$ closes, resulting in the emergence of an isolated shell-like region. The black lines in Figure \ref{fig:trajectories_for_Ne}(e) delineate this region, while the pink lines represent the dynamics associated with the near-field attractive electrostatic interaction. Trajectories shown by the green lines that originate outside this shell will diverge toward the far field upon forward integration. In conclusion, this case demonstrates that no trajectory originating from the far field can cross the repulsive barrier and reach the collision sphere, consistent with the collision suppression criterion $N_e>(N_e)_c$.

A dynamical system interpretation of Figure \ref{fig:trajectories_for_Ne} can be derived by analyzing the motion of particle pairs as an autonomous two-dimensional system described by the following equations:
\begin{subeqnarray}
    \dot{r} & = & v_r=-r(1-A)(3\cos^2\theta-1)+N_eGf_e, \label{rdot_equation} \\[3pt]
    \dot{\theta} & = & \frac{v_\theta}{r} = 3(1-B)\sin\theta\cos\theta,  \label{thetadot_equation}
\end{subeqnarray}
for $N_v=0$. To find the equilibrium points, we require that $v_r=0$ and $v_\theta=0$. The latter condition restricts the possible equilibrium points to the compressional axis, $\theta=0, \pi$, and the equatorial plane, $\theta=\pi/2$.

On the equatorial plane, the equilibrium condition becomes
\begin{equation}
r(1-A)+N_eGf_e=0. \label{equatorial plane_fixed_points}
\end{equation}
For the parameter combination considered in Figure \ref{fig:trajectories_for_Ne}, the electrostatic interaction is attractive in the near field and repulsive in the far field. This near-field attraction can balance the outward flow-induced radial motion in the equatorial plane, resulting in an equatorial equilibrium at $r=r_s$. Linearization about this equilibrium yields two eigenvalues:
\begin{equation}
\lambda_r=\left.\frac{\partial}{\partial r} \left[r(1-A)+N_eGf_e\right] \right|_{r=r_s},
\qquad 
\lambda_{\theta} = \left.\frac{\partial \dot{\theta}}{\partial \theta} \right|_{\theta=\pi/2} = -3(1-B(r_s)). \label{lambda_r_and theta}
\end{equation}
For the equilibrium shown in the present phase portraits, $\lambda_r > 0$ while $\lambda_\theta < 0$, identifying it as a saddle point: radial perturbations are unstable because trajectories just inside the equilibrium are driven towards contact, whereas those just outside it move outward; conversely, angular perturbations are restoring towards $\theta=\pi/2$. The portion of the limiting colliding trajectory can be interpreted as a branch of the stable manifold of this saddle point, which separates colliding from non-colliding initial conditions. The inward part of this separatrix forms part of the unstable manifold leading towards the collision sphere.

At $N_e=1$, this equatorial saddle point lies relatively close to the collision sphere. As $N_e$ increases to $25$ and $29$, this equatorial saddle point moves away from the collision sphere, and its stable manifold becomes increasingly deformed by the intermediate-range electrostatic repulsion. This results in a progressively narrower neck observable along the compressional direction. For $N_e=25$ and $29$, both of which satisfy $N_e < (N_e)_c \approx 30.4$, no equilibrium exists on the compressional axis. Nonetheless, the radial velocity along this axis diminishes considerably over an intermediate range of separations as $N_e$ approaches $(N_e)_c$, giving rise to a dynamical bottleneck through which colliding trajectories can still pass. 

At $N_e=(N_e)_c$, the equilibrium condition along the compressional axis is governed by the equation
\begin{equation}
-2r(1-A)+N_eGf_e=0. \label{compressional_axis_equilibrium}
\end{equation}
This equation first yields a double root at $r=r_c$ (see the inset of Figure \ref{fig:vr_vs_xi}), satisfying simultaneously
\begin{equation}
v_r(r_c,0)=0
\quad \text{and} \quad
\left.\frac{\partial v_r}{\partial r}\right|_{r=r_c}=0, \label{criticality_condition}
\end{equation}
which is precisely the criticality condition used to define $(N_e)_c$.

For $N_e = 40 > (N_e)_c$, the degenerate equilibrium splits into two axial fixed points on each compressional half-axis. Linearizing the system about an axial fixed point $r=r_a$, $\theta=0$, gives
\begin{equation}
\lambda_r=\left.\frac{\partial}{\partial r} \left[-2r(1-A)+N_eGf_e\right] \right|_{r=r_a}, \qquad
\lambda_{\theta} = \left.\frac{\partial \dot{\theta}}{\partial \theta} \right|_{\theta=0} = 3\left(1-B(r_a)\right), \label{lambda_r_and_theta_axial}
\end{equation}
where $\lambda_\theta$ is positive because the background flow drives particle pairs away from the compressional axis. At the outer fixed point ($r_a \approx 6$), $\lambda_r<0$, so the fixed point is a saddle, wherein radial perturbations are restorative while transverse perturbations are unstable. Conversely, at the inner fixed point ($r_a \approx 4.2$), $\lambda_r>0$, so it is an unstable node.

The axial fixed points occurring for $N_e>(N_e)_c$ notably do not eliminate the pre-existing equatorial saddle points. For the parameters considered here, the equatorial saddle point persists at $N_e=40$ and therefore coexists with the newly created axial fixed points. Consequently, the transition across $(N_e)_c$ modifies the structure of equilibria specifically along the compressional axis, while the equatorial saddle points continue to emerge from the interplay between the outward flow-induced motion and the attraction induced by near-field polarization. In the full axisymmetric configuration, these equatorial saddle points correspond to a ring of equivalent equilibria, whose intersections with the meridional plane appear as symmetric saddle points. It is important to note, however, that the existence of these points is not universal; it requires sufficiently strong near-field electrostatic attraction to balance the outward motion in the equatorial plane.

The invariant manifolds of the axial fixed points reorganize the global trajectory topology and prevent trajectories arriving from the far field from penetrating into the near-field attractive region. In particular, the change in global topology is governed by the stable manifold of the equatorial saddle, which constitutes the limiting colliding trajectory and therefore separates colliding from non-colliding initial conditions. For $N_e<(N_e)_c$, this manifold extends to the far field, thereby allowing a finite set of far-field trajectories to reach the collision sphere. In contrast, for $N_e>(N_e)_c$, the manifold originates at the inner axial node rather than in the far field. Consequently, the near-contact attractive region becomes dynamically separated from particles approaching from infinity, and trajectories leading to contact are confined to the bounded shell-like region, as shown in Figure \ref{fig:trajectories_for_Ne}(e). Thus, the transition from $N_e=29$ to $N_e=40$ can be interpreted as a saddle-node bifurcation on the compressional axis. Just below criticality, there is a narrow dynamical bottleneck, whereas just above criticality, newly created saddle points and their separatrices disconnect the near-contact collision region from the far field, thereby causing the collision efficiency to vanish.

The trajectories shown in Figure \ref{fig:trajectories_for_Ne}(b-e) correspond to size and charge ratios for which the electrostatic interaction becomes attractive in the near field. When the electrostatic interaction remains repulsive at all separations, the near-field attractive region and the associated equatorial saddle points are absent. Nevertheless, the axial mechanism of collision suppression remains qualitatively similar: increasing $N_e$ progressively reduces the upstream interception area, and, for $N_e>(N_e)_c$, the electrostatic barrier prevents trajectories originating in the far field from reaching the collision sphere. Therefore, we expect the collision efficiency to decrease monotonically with increasing $N_e$ and vanish once the imposed compressional flow can no longer overcome the repulsive barrier.

The same dynamical system framework remains applicable for finite $N_v$, because the van der Waals interactions modify only the radial relative velocity equation. Consequently, the equilibrium locations remain confined to the compressional axis and the equatorial plane, although the additional attractive interaction modifies their radial positions. The van der Waals attraction leads to an increase in $(N_e)_c$, necessitating a stronger electrostatic repulsion to counteract both the imposed compressional motion and the attractive van der Waals contribution. Similarly, the equatorial fixed points shift as $N_v$ varies. Thus, while the same dynamical system analysis is applicable for finite $N_v$, the locations and stability of the fixed points, the bifurcation threshold, and, in cases of sufficiently strong van der Waals attraction, the detailed phase-space topology may differ from the case where $N_v = 0$.

The limiting case $N_e=\infty$, shown in Figure \ref{fig:trajectories_for_Ne}(f), represents a situation wherein particle motion is governed solely by electrostatic interactions, with no influence from the background flow. In this scenario, the dimensionless radial relative velocity can be expressed as:
\begin{equation}
    v_r^{(e)} =
    \left(\frac{1+\kappa}{\kappa}\right)Gf_e .
    \label{Eq:Ne_infty_rad_vel}
\end{equation}
In the above equation, the relative velocity is scaled by $q_1^2/(4\pi\epsilon_0 a_1^2)/
[6 \pi \mu_f a_1]$, which is the characteristic relative velocity associated with purely electrostatically driven relative motion.

Since the electrostatic force acts purely in the radial direction, the relative trajectories do not depend on the angular coordinates $\theta$ and $\phi$. In Figure \ref{fig:trajectories_for_Ne}(f), the thick black circle delineates the boundary at which $f_e=0$. Inside this circle, polarization-induced attraction leads to radially inward trajectories, whereas outside it, the electrostatic interactions are repulsive, directing trajectories radially outward. Consequently, particles initially located outside this critical separation cannot reach the collision sphere in the absence of an imposed flow. This limiting case highlights the role of the compressional flow at finite $N_e$, especially for $N_e<(N_e)_c$, where the flow facilitates particle pairs to traverse the intermediate repulsive region and enter the near-field attractive regime, where collision can occur.

\subsection{Non-monotonic variation of the collision efficiency} \label{subsec:E12_Ne}

The preceding analysis demonstrates that changes in the topology of the relative trajectories directly modify the upstream interception area and, consequently, the collision rate. We therefore examine the dependence of $E_{12}$ on the key parameters $\kappa$, $\beta$, $N_e$, $Kn$, and $N_v$, which collectively govern the collision dynamics. In the limit $N_e \rightarrow 0$, electrostatic interactions become negligible, and $E_{12}$ approaches a value determined by the combined effects of the imposed compressional flow, non-continuum lubrication interactions, and van der Waals forces. In this flow-dominated regime, the collision efficiency increases with increasing $\kappa$. For small $\kappa$, the large disparity in particle sizes causes the smaller particle to follow the imposed flow streamlines more closely and to move around the larger particle rather than collide with it. Consequently, collisions at small $\kappa$ are restricted to particles approaching along streamlines that pass sufficiently close to the larger particle. As $\kappa$ increases, this restriction becomes less pronounced, resulting in a larger upstream interception area and, hence, a higher collision efficiency. As $N_e$ increases, electrostatic interactions play an increasingly prominent role, thereby modifying the collision dynamics in ways that can either enhance or attenuate $E_{12}$, contingent upon the values of $\kappa$ and $\beta$. Nevertheless, for sufficiently strong electrostatic interactions, repulsion ultimately prevents particles from reaching the collision sphere, and $E_{12}$ decreases to zero as $N_e$ approaches $(N_e)_c$.

\begin{figure}
  \centerline{\includegraphics[width=1.0\textwidth]{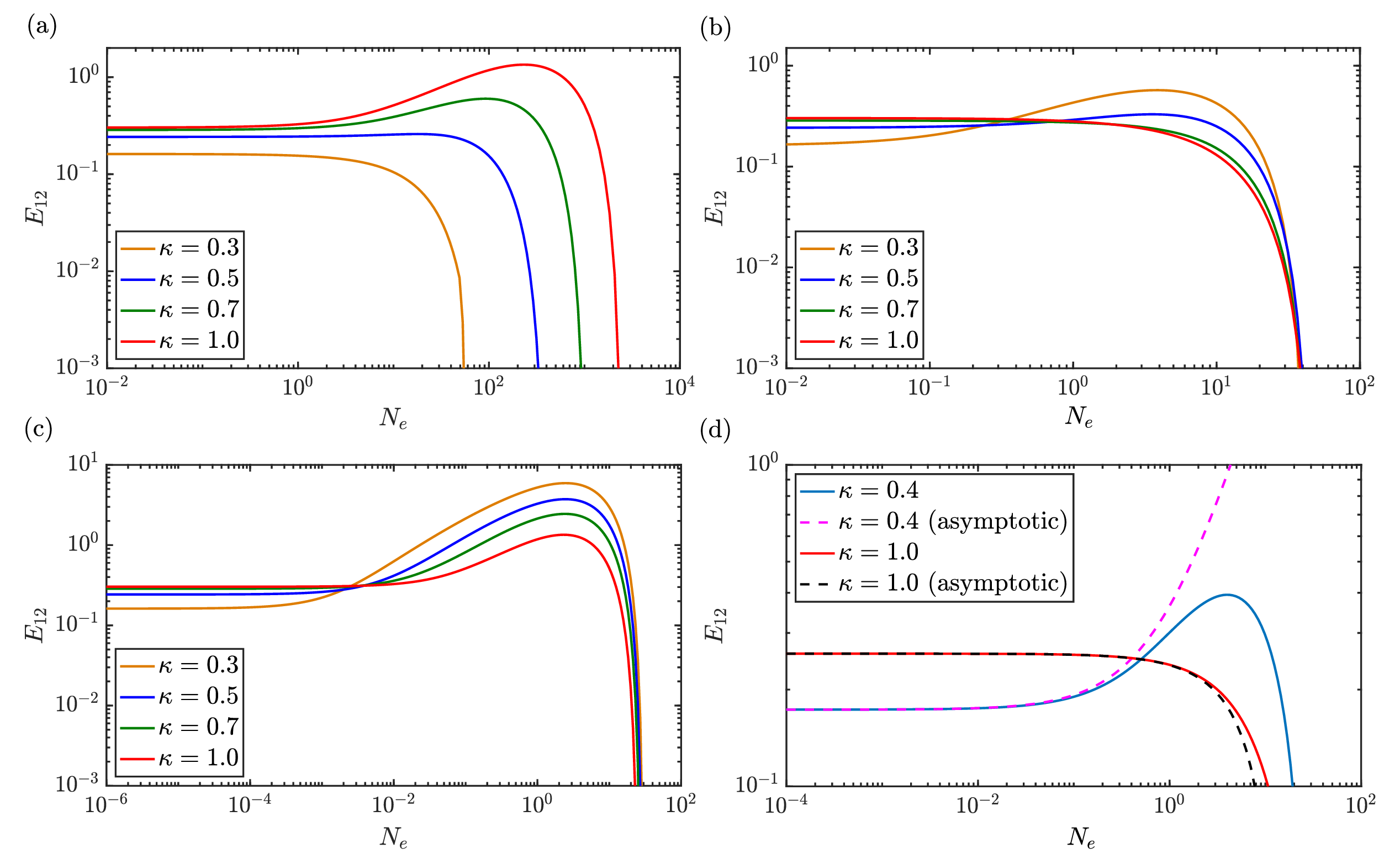}}
  \caption{(a-c) Collision efficiencies as a function of $N_e$ for $Kn=10^{-2}$, $\kappa = 0.3, 0.5, 0.7, 1.0$, $N_L = 500$, $N_v = 10^{-3}$ and $k_1=k_2=80$ when (a) $\beta = 10^{-1}$, (b) $\beta = 1$, and (c) $\beta = 10$. (d) Collision efficiency as a function of $N_e$ for $Kn=10^{-2}$, $k_1=k_2=80$, $\beta=1$, and $\kappa=0.4$ and $1.0$ in the absence of van der Waals interactions ($N_v=0$). The dashed lines in (d) denote the corresponding asymptotic predictions from Eq. \eqref{eq:E12_small_Ne}.}
\label{Collision_efficiency_vs_Ne_with_vdW}
\end{figure}

Figure \ref{Collision_efficiency_vs_Ne_with_vdW}(a-c) shows the variation of $E_{12}$ with $N_e$ for $\kappa=0.3$, $0.5$, $0.7$, and $1.0$, with $\beta=10^{-1}$, $1$, and $10$ in panels (a-c), respectively, for $Kn=10^{-2}$ and $N_v=10^{-3}$. For the small charge ratio $\beta=10^{-1}$ [Figure \ref{Collision_efficiency_vs_Ne_with_vdW}(a)], the near-field electrostatic attraction becomes stronger with increasing $\kappa$ for $\kappa \gtrsim 0.32$. Consequently, the impact of near-field electrostatic attraction on $E_{12}$ becomes increasingly significant as the size ratio extends beyond this critical value. For $\kappa=0.3$, electrostatic interactions remain repulsive at all separations, causing $E_{12}$ to decrease monotonically from its $N_e=0$ value as $N_e$ increases. However, for larger size ratios $\kappa=0.5$, $0.7$, and $1.0$, the presence of strong near-field attraction results in a non-monotonic variation of $E_{12}$. In these cases, the collision efficiency exhibits an initial increase above its value at $N_e=0$, followed by a subsequent decline as the electrostatic repulsive barrier becomes sufficiently strong. Ultimately, the collision efficiency vanishes as $N_e$ approaches the corresponding critical value $(N_e)_c$. It is noteworthy that for $\beta=10^{-1}$ and $\kappa$ values ranging from $0.3$ to $1$, the electrostatic effects favour the variation of collision efficiency with $\kappa$ in the $N_e \rightarrow 0$ limit. As a consequence, the curves representing different $\kappa$ values in Figure \ref{Collision_efficiency_vs_Ne_with_vdW}(a) do not intersect with one another.

A qualitatively different dependence on $\kappa$ emerges for $\beta=1$ [Figure \ref{Collision_efficiency_vs_Ne_with_vdW}(b)]. In the flow-dominated regime, the collision efficiency increases with increasing $\kappa$. As $N_e$ increases, however, this trend reverses because the near-field electrostatic attraction becomes progressively weaker with increasing size ratio when $\beta=1$. Notably, for $\kappa=1$, the near-field electrostatic interaction remains repulsive. Consequently, the curves corresponding to different size ratios intersect at intermediate values of $N_e$, reflecting the contrasting dependence of the flow-driven and electrostatically modified collision dynamics on $\kappa$. Consistent with the presence of near-field electrostatic attraction, the collision efficiencies for $\kappa=0.3$, $0.5$, and $0.7$ exceed their respective $N_e=0$ values over an intermediate range of $N_e$ before decreasing sharply as $N_e$ approaches the corresponding $(N_e)_c$. In contrast, no such enhancement occurs for $\kappa=1$, owing to the repulsive nature of the electrostatic interaction.

For $\beta=10$ [Figure \ref{Collision_efficiency_vs_Ne_with_vdW}(c)], the variation of $E_{12}$ with $N_e$ for different values of $\kappa$ is qualitatively similar to that observed for $\beta=1$. However, the underlying electrostatic interactions that drive this behaviour differ in two important respects. First, the near-field electrostatic interaction is attractive for all values of $\kappa$ considered. Second, the magnitude of this attraction is substantially greater than that for $\beta=1$. These characteristics result in a more pronounced electrostatic enhancement of particle collisions, resulting in $E_{12}$ exceeding its corresponding value at $N_e=0$ over an intermediate range of $N_e$ for all values of $\kappa$. Notably, for smaller values of $\kappa$, this enhancement can be as much as an order of magnitude relative to the corresponding collision efficiency at $N_e=0$.

\subsection{Weak electrostatic interactions} \label{subsec:weak_Ne}

Figure \ref{Collision_efficiency_vs_Ne_with_vdW}(a-c) shows that weak charges can either raise or lower the collision efficiency, depending on $\kappa$ and $\beta$. To determine which separations control the sign and magnitude of this initial change, we treat the electrostatic contribution as a regular perturbation of the purely hydrodynamic relative trajectories for $N_e\ll 1$. We set $N_v=0$, so that contact is enabled by non-continuum lubrication alone, and write
\begin{equation}
    \theta(r)=\theta_0(r)+N_e\theta_1(r)+O(N_e^2),
    \label{eq:theta_expansion_small_Ne}
\end{equation}
where $\theta_0$ denotes the leading-order trajectory in the absence of electrostatic interactions and $\theta_1$ represents the first-order electrostatic correction.

The limiting colliding trajectory satisfies the condition $v_r=0$ at the collision surface. At leading order, this gives
\begin{equation}
    \theta_0(2) = \cos^{-1}\left(\frac{1}{\sqrt{3}}\right), \label{eq:theta0_boundary_small_Ne}
\end{equation}
whereas the first-order condition yields
\begin{equation}
    \theta_1(2) = -\left. \frac{Gf_e} {6r(1-A)\sin\theta_0\cos\theta_0} \right|_{r\rightarrow 2^+} \equiv C_1. \label{eq:theta1_boundary_small_Ne}
\end{equation}
The limit $r\rightarrow2^+$ is understood because both $1-A$ and $G$ vanish at exact contact, while their ratio remains finite. At leading order, the limiting colliding trajectory is the purely hydrodynamic trajectory (\ref{relative_trajectory_expression_only_flow}) with $C=2/(3\sqrt{3})$. Defining $I(r)= 2 \varphi(r)/(3\sqrt{3})$, it satisfies
\begin{equation}
    \sin^2\theta_0(r)\cos\theta_0(r)=I(r),
    \label{eq:theta0_implicit_small_Ne}
\end{equation}
and selecting the root corresponding to the limiting colliding trajectory yields
\begin{equation}
    \theta_0(r) = \cos^{-1} \left\{\frac{2}{\sqrt{3}}\cos\left[\frac{1}{3}\cos^{-1}
    \left(-\frac{3\sqrt{3}}{2}I(r)\right)\right]\right\}. \label{eq:theta0_solution_small_Ne}
\end{equation}
This solution satisfies $\theta_0(2)=\cos^{-1}(1/\sqrt{3})$ and approaches $\theta_0\rightarrow0$ as $r\rightarrow\infty$, as required for the upstream limiting trajectory.

The first-order correction to the collision efficiency can be written as a single quadrature along this trajectory. Let $s(\theta)=\sin^2\theta\cos\theta$, so that $C=s(\theta)/\varphi(r)$ is the constant of (\ref{relative_trajectory_expression_only_flow}) when $N_e=0$. For $N_v=0$, the trajectory equation (\ref{final_trajectory_equation}) gives, exactly,
\begin{equation}
    \frac{\mathrm{d}}{\mathrm{d}r}\ln\left[\frac{s(\theta)}{\varphi(r)}\right] = \frac{3(1-B)}{r(1-A)}\,\frac{N_e G f_e}{v_r},
    \label{eq:dlnC_small_Ne}
\end{equation}
so that $C$ changes along a trajectory only through the electrostatic contribution to $v_r$. The collision efficiency is proportional to the far-field value of $C$ on the limiting colliding trajectory. The shift (\ref{eq:theta1_boundary_small_Ne}) of its contact point changes the value of $C$ at contact only at $O(N_e^2)$, because $s(\theta)$ is stationary at $\theta_0(2)=\cos^{-1}(1/\sqrt{3})$. Integrating (\ref{eq:dlnC_small_Ne}) along the leading-order trajectory (\ref{eq:theta0_solution_small_Ne}), on which $v_r=-r(1-A)(3\cos^2\theta_0-1)$, gives
\begin{equation}
    E_{12}=E_{12}^{(0)}\left[1-N_e\int_2^\infty W(r)\,G(r)\,f_e(r)\,\mathrm{d}r\right]+O(N_e^2),
    \label{eq:E12_small_Ne}
\end{equation}
where $E_{12}^{(0)}$ is the collision efficiency of uncharged, hydrodynamically interacting spheres given by (\ref{Analytical_expression_collision_efficiency_only_flow}) and
\begin{equation}
    W(r)=\frac{3(1-B)}{r^2(1-A)^2\,(3\cos^2\theta_0-1)}.
    \label{eq:W_small_Ne}
\end{equation}
The weight $W$ is positive, so that electrostatic repulsion lowers the collision efficiency at first order and attraction raises it. Although $W(r)$ becomes singular as $r \to 2^+$, this singularity is integrable. In this limit, $G/(1-A)$ remains finite and $3\cos^2\theta_0-1 = O\!\left(J^{1/2}\right)$, where $J(r)=\int_2^r \mathrm{d}r'/(1-A(r'))$. Owing to the weaker divergence of the non-continuum lubrication resistance, $J(r)$ converges, consistent with finite-time contact. Consequently, the integrand in \eqref{eq:E12_small_Ne} behaves as $O\!\left(J^{-1/2}\,\mathrm{d}J\right)$, whose integral is finite. The leading correction is thus a weighted average of the electrostatic force along the limiting trajectory, with the weight concentrated in the lubrication region; this is why near-contact attraction can raise $E_{12}$ even when the interaction is repulsive at larger separations. The expansion requires the electrostatic contribution to $v_r$ at contact to remain small compared with the straining contribution, $N_e\,|Gf_e/[r(1-A)]|_{r\rightarrow2^+}\ll1$, rather than merely $N_e\ll1$. When the near-contact force is attractive and this ratio reaches unity, the inflow region on the collision sphere extends to the equator and the equatorial saddle point discussed in \S \ref{subsec:topology} emerges from the collision sphere; beyond this point, the limiting colliding trajectory is the stable manifold of that saddle point, which a regular expansion about the purely hydrodynamic trajectory cannot describe.

Figure \ref{Collision_efficiency_vs_Ne_with_vdW}(d) compares the numerically calculated collision efficiencies with the asymptotic prediction (\ref{eq:E12_small_Ne}) for $Kn=10^{-2}$, $\beta=1$, $N_v=0$, and $\kappa=0.4$ and $1.0$. In the limit $N_e \ll 1$, the asymptotic results closely match the numerical solutions for both size ratios, including the increase of $E_{12}$ with $N_e$ for $\kappa=0.4$, for which the near-contact interaction is attractive, and its decrease for $\kappa=1.0$, for which it is repulsive. The asymptotic and numerical results separate at $N_e \approx 10^{-1}$ for $\kappa=0.4$ but only at $N_e \approx 1$ for $\kappa=1.0$, consistent with the condition stated above; a full trajectory calculation is therefore required for moderate and large values of $N_e$.

\subsection{Effects of non-continuum lubrication, van der Waals attraction and charge ratio} \label{subsec:Kn_Nv_beta}

\begin{figure}
  \centerline{\includegraphics[width=1.0\textwidth]{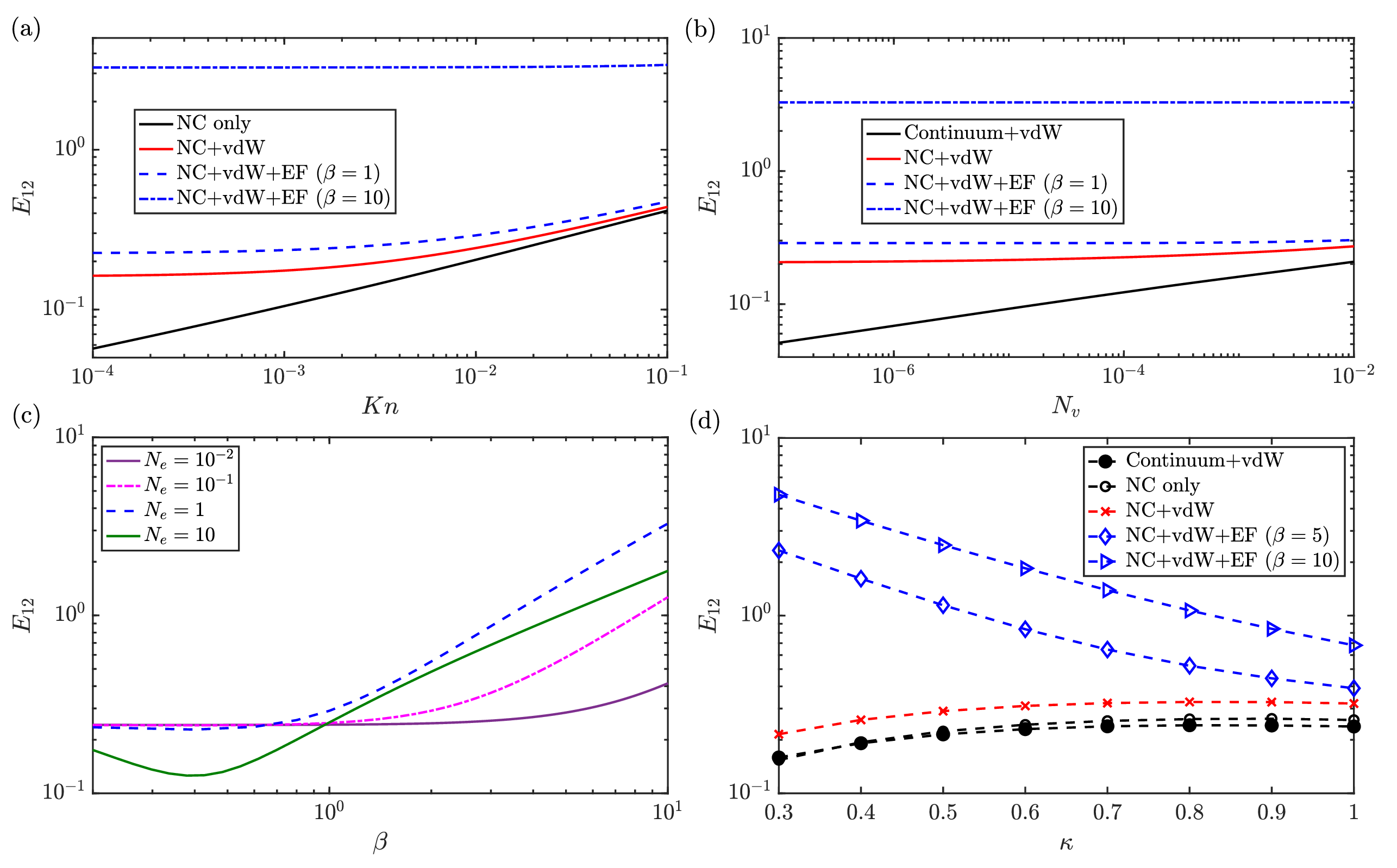}}
  \caption{(a) Collision efficiency, $E_{12}$, as a function of $Kn$ for non-continuum hydrodynamic interactions alone (NC only; solid black line), non-continuum hydrodynamic and van der Waals interactions (NC+vdW; solid red line), and non-continuum hydrodynamic, van der Waals, and electrostatic interactions (NC+vdW+EF) for $\beta=1$ (dotted blue line) and $\beta=10$ (dash-dotted blue line), with $\kappa=0.5$, $N_e=1$, $N_L=250$, and $N_v=10^{-3}$. (b) $E_{12}$ as a function of $N_v$ for continuum hydrodynamic and van der Waals interactions (Continuum+vdW; solid black line), NC+vdW (solid red line), and NC+vdW+EF for $\beta=1$ (dotted blue line) and $\beta=10$ (dash-dotted blue line), with $\kappa=0.5$, $Kn=10^{-2}$, $N_e=1$, and $N_L=250$. (c) $E_{12}$ as a function of the charge ratio $\beta$ for $N_e=10^{-2}$, $10^{-1}$, $1$, and $10$, with $\kappa=0.5$, $Kn=10^{-2}$, $N_L=250$, and $N_v=10^{-3}$. (d) $E_{12}$ as a function of the size ratio $\kappa$ for water droplets in air with $a_1=10\,\text{\textmu m}$ and $q_1=200e$, for $\beta=5$ and $10$, with NC+vdW+EF. For comparison, the collision efficiencies of uncharged droplets in a uniaxial compressional flow reported by \citet{dhanasekaran2021collision} are also shown.}
\label{fig:Collision_efficiency_with_Kn_Nv_beta_and_kappa}
\end{figure}

We now consider how non-continuum lubrication effects, measured by the Knudsen number $Kn$, modify the collision efficiency in the presence of van der Waals and electrostatic interactions. Figure \ref{fig:Collision_efficiency_with_Kn_Nv_beta_and_kappa}(a) shows the variation of $E_{12}$ with $Kn$ for $\kappa=0.5$, $N_e=1$, $N_L=250$, and $N_v=10^{-3}$. For comparison, we also show results without electrostatic interactions. With non-continuum (NC) hydrodynamic interactions alone, $E_{12}$, obtained by evaluating the integral in (\ref{Analytical_expression_collision_efficiency_only_flow}), decreases monotonically with decreasing $Kn$ and approaches zero as $Kn \rightarrow 0$, since the relative thickness of the non-continuum lubrication region progressively diminishes. The inclusion of van der Waals (vdW) attraction changes this asymptotic behaviour: for the NC+vdW case, $E_{12}$ approaches a finite value as $Kn \rightarrow 0$, corresponding to collisions driven by the imposed compressional flow and vdW attraction in the presence of continuum hydrodynamic interactions. Accounting for electrostatic interactions arising from particle surface charges further enhances $E_{12}$ when polarization gives rise to an attractive near-field interaction. For $\beta=1$, the dependence of $E_{12}$ on $Kn$ remains qualitatively similar to that for NC+vdW. However, for $\beta=10$, the substantially stronger near-field electrostatic attraction dominates the collision dynamics, rendering $E_{12}$ nearly independent of $Kn$ over the range considered.

The sensitivity of the collision efficiency to van der Waals interactions is quantified by varying $N_v$ from $10^{-7}$ to $10^{-2}$. Figure \ref{fig:Collision_efficiency_with_Kn_Nv_beta_and_kappa}(b) shows the variation of $E_{12}$ with $N_v$ for $\kappa=0.5$, $N_e=1$, and $N_L=250$. In the absence of electrostatic interactions, decreasing $N_v$ reduces $E_{12}$, with the reduction being particularly pronounced for continuum hydrodynamics. For non-continuum lubrication interactions ($Kn=10^{-2}$), the decrease is more gradual, and $E_{12}$ approaches the value obtained in the absence of van der Waals interactions. Accounting for electrostatic interactions enhances $E_{12}$ when the near-field interaction is attractive, as for $\beta=1$ and $10$. For $\beta=10$, the near-field electrostatic attraction is sufficiently strong that $E_{12}$ becomes nearly insensitive to variations in the strength of the van der Waals interaction, implying that electrostatic forces dominate over other collision-inducing mechanisms.

To examine the influence of the relative particle charges on the collision dynamics, in Figure \ref{fig:Collision_efficiency_with_Kn_Nv_beta_and_kappa}(c), we present $E_{12}$ as a function of the charge ratio $\beta$ for four different values of $N_e$. For weak electrostatic interactions ($N_e=10^{-2}$), $E_{12}$ depends only weakly on $\beta$, indicating predominantly flow-driven collision dynamics. As $N_e$ increases, the dependence on $\beta$ becomes progressively stronger. For $N_e=1$ and $10$, $E_{12}$ initially decreases with increasing $\beta$, reflecting the suppression of collisions by electrostatic repulsion, before increasing as polarization-induced attraction becomes increasingly important. This competition is particularly pronounced for $N_e=10$, for which $E_{12}$ exhibits a distinct minimum at small $\beta$. At larger $\beta$, the rapid increase in $E_{12}$ demonstrates the increasing influence of near-field electrostatic attraction on the collision dynamics. For any prescribed $\beta$, the dependence of $E_{12}$ on $N_e$ is consistent with the trends discussed earlier.

Finally, we estimate the collision efficiency of like-charged droplet pairs with unequal sizes suspended in air. Figure \ref{fig:Collision_efficiency_with_Kn_Nv_beta_and_kappa}(d) shows $E_{12}$ as a function of $\kappa$ for water droplets in air with $a_1=10$ \textmu m, $q_1=200e$, $\dot{\gamma}=25$ s$^{-1}$ and $\lambda_0=0.1$ \textmu m. For these conditions, the relevant dimensionless parameters vary with size ratio as $Kn=0.02/(1+\kappa)$, $N_e = 0.22/\kappa$, $N_L = 200\pi(1+\kappa)$, and $N_v = 1.9\times10^{-2}/[\kappa(1+\kappa)]$. The charge ratios $\beta=5$ and $10$, for which the smaller droplet carries the larger charge, depart from the empirical relations $\beta=\kappa^{1.3}$ and $\beta=\kappa^{2}$ for natural cloud droplets used in \S \ref{subsec:Critical_Ne}, and are more representative of artificially charged droplets \citep{khain2004rain}. In the absence of electrostatic interactions, $E_{12}$ increases with $\kappa$, consistent with the findings of \citet{dhanasekaran2021collision}. Electrostatic interactions qualitatively reverse the dependence of $E_{12}$ on $\kappa$ observed for uncharged droplets: for both $\beta=5$ and $10$, the collision efficiency decreases monotonically with increasing $\kappa$. This behaviour arises from the stronger polarization-induced attraction experienced by droplet pairs with greater size disparity at these charge ratios, which facilitates particle approach and thereby counteracts the comparatively low collision efficiency associated with hydrodynamic interactions alone at small $\kappa$. Furthermore, $E_{12}$ is systematically larger for $\beta=10$ than for $\beta=5$ over the range of $\kappa$ considered, reflecting that the near-field electrostatic attraction force becomes increasingly pronounced with increasing charge ratio. These results demonstrate that, under the conditions considered, electrostatic interactions can substantially increase the collision efficiency of like-charged droplet pairs with large disparities in both size and charge compared with their uncharged counterparts.

\section{Conclusions}
\label{sec: Conclusion}

We have investigated the collision dynamics of like-charged dielectric particles subjected to a uniaxial compressional flow, accounting for the coupled effects of hydrodynamic and electrostatic interactions. The formulation incorporates non-continuum lubrication effects, retarded van der Waals attraction, and the complete electrostatic interaction between finite-sized dielectric spheres. This approach effectively addresses the mutual polarization effects, which become increasingly important as the particles approach one another. Our analysis of charged tracers elucidated the competing influences of Coulomb repulsion and polarization-induced attraction on collision dynamics. For finite-sized, hydrodynamically interacting particles, we identified a critical value of the dimensionless electrostatic parameter, $(N_e)_c$, which signifies the transition from collision to collision suppression. When the parameter exceeds this threshold, the repulsive electrostatic barrier cannot be overcome by the flow-induced inward relative motion, thereby precluding collisions between particles approaching from the far field. The value of $(N_e)_c$ depends strongly on the particle size ratio $\kappa$ and charge ratio $\beta$, reflecting the sensitivity of the finite-dielectric electrostatic interaction to particle size and charge asymmetry.

The representative pair trajectories shown in Figure \ref{fig:trajectories_for_Ne} reveal how changes in trajectory topology govern the transition from collision to collision suppression. For $N_e < (N_e)_c$, the inward relative motion induced by the compressional flow is sufficient to overcome the intermediate electrostatic repulsive barrier, allowing the particles to reach the near-field region where, when present, polarization-induced attraction further promotes their approach to contact. As $N_e$ approaches $(N_e)_c$, the upstream interception area associated with colliding trajectories progressively contracts and the collision efficiency decreases. For $N_e>(N_e)_c$, trajectories originating in the far field can no longer reach the collision surface, and the collision efficiency vanishes. We have shown that electrostatic interactions can either enhance or inhibit collisions depending on $\kappa$, $\beta$, and $N_e$. When polarization-induced attraction becomes sufficiently strong, it can substantially enhance the collision efficiency and diminish the influence of non-continuum lubrication and van der Waals interactions on the collision dynamics. In contrast, sufficiently strong electrostatic repulsion prevents particles from coming into contact, causing collision efficiency to vanish once $N_e$ exceeds $(N_e)_c$.

Two analytical results support this picture. The threshold $(N_e)_c$ corresponds to a saddle-node bifurcation on the compressional axis, along which the inward straining is strongest. When the near-contact interaction is attractive, the limiting colliding trajectory is the stable manifold of an equatorial saddle point; above $(N_e)_c$ this manifold originates at the newly created axial node rather than in the far field, so that the trajectories reaching contact occupy a bounded region disconnected from infinity and the collision efficiency vanishes. For weak electrostatic interactions, a regular perturbation of the limiting colliding trajectory expresses the $O(N_e)$ change in the collision efficiency as a weighted average of the electrostatic force along that trajectory, with the weight concentrated in the lubrication region, so that the near-contact interaction weighs most heavily in deciding whether weak charges enhance or suppress collisions. This prediction agrees closely with the full trajectory calculations.

As discussed in \S \ref{sec:Introduction}, the present study is relevant to electrically active particulate flows encountered in both atmospheric and geophysical settings. In particular, electrostatic interactions may influence droplet collisions in strongly electrified clouds, and particle aggregation in the `dirty thunderstorms' associated with explosive volcanic eruptions, where ash particles can acquire substantial electrical charge. The present results establish the underlying particle scale physics by showing how the competition between hydrodynamic and electrostatic forces determines whether charged particles collide. Although these processes occur within complex turbulent environments, resolving their interactions at the particle level is a necessary step towards constructing physically based collision kernels for charged particles in turbulence.

The present trajectory-based framework can be extended to turbulent flows by approximating the velocity gradient experienced by a sub-Kolmogorov particle pair as a linearly varying stochastic field. This approximation will allow us to integrate the equations governing the relative motion of finite-size charged particles over many statistically independent realizations of the velocity-gradient tensor. Following the Monte Carlo integration approach developed by \citet{dhanasekaran2021tubulent}, the turbulent collision rate can be obtained by ensemble averaging the collision flux over these realizations. This extension would therefore quantify how temporal fluctuations in the local strain and rotation rate tensors modify the collision efficiency calculated for the steady compressional flow.

Future work should additionally incorporate particle inertia and gravitational settling, both of which can alter relative velocities and, in turbulent flows, the spatial distribution of particle pairs. In particular, particle inertia can lead to preferential concentration and caustics, which cannot be captured by the deterministic linear-flow formulation considered here. An approach combining these effects with finite-size hydrodynamic and electrostatic interactions would enable the relative importance of turbulence, gravity, and electrostatic force to be quantified systematically across relevant particle sizes and charges. The resulting collision kernels could subsequently be used in population-balance or particle-based models to examine the consequences of electrostatic forces on droplet growth and precipitation initiation in thunderstorms, and on ash aggregation and fallout in volcanic `dirty thunderstorms'.

\vspace{5mm}
\textbf{Funding.} P.P. would like to acknowledge financial support from the Prime Minister’s Research Fellows (PMRF) scheme, Ministry of Education, Government of India (Project Number: SB22230184AMPMRF008746). A.R. acknowledges the financial support received from the Anusandhan National Research Foundation (ANRF), Government of India, through the Advanced Research Grant (ANRF/ARG/2025/010666/ENS).

\vspace{5mm}
\textbf{Declaration of interests.} The authors report no conflict of interest.

\vspace{5mm}
\textbf{Data availability statement.} The data and code supporting the findings of this study are available from the corresponding author upon reasonable request.

\bibliographystyle{jfm}
\bibliography{jfm}

\end{document}